\documentclass[a4paper,fleqn,usenatbib]{mnras}

\usepackage{newtxtext,newtxmath}

\usepackage[T1]{fontenc}
\usepackage{ae,aecompl}

\usepackage{graphicx}	
\usepackage{amsmath}	
\usepackage{amssymb}	
\usepackage{pdfpages}

\title[Sulphur chemistry in hot cores]{A new look at sulphur chemistry in hot cores and corinos}

\author[T. Vidal et al.]{
Thomas H. G. Vidal,$^{1}$\thanks{E-mail: thomas.vidal@u-bordeaux.fr}
and Valentine Wakelam$^{1}$
\\
$^{1}$Laboratoire d'astrophysique de Bordeaux, Univ. Bordeaux, CNRS, B18N, allée Geoffroy Saint-Hilaire, 33615 Pessac, France\\
}

\date{Accepted XXX. Received YYY; in original form ZZZ}

\pubyear{2017}

\begin{document}
\label{firstpage}
\pagerange{\pageref{firstpage}--\pageref{lastpage}}
\maketitle

\begin{abstract}
Sulphur-bearing species are often used to probe the evolution of hot cores since their abundances are particularly sensitive to physical and chemical variations. However, the chemistry of sulphur is not well understood in these regions, notably because observations of several hot cores displayed a large variety of sulphur compositions, and because the reservoir of sulphur in dense clouds, in which hot cores form, is still poorly constrained. In order to help disentangled its complexity, we present a fresh comprehensive review of sulphur chemistry in hot cores along with a study of its sensibility to temperature and pre-collapse chemical composition. In parallel, we analyse the discrepencies that result from the use of two different types of models (static and dynamic) to highlight the sensitivity to the choice of model to be used in astrochemical studies. Our results show that the pre-collapse chemical composition is a critical parameter for sulphur chemistry in hot cores and could explain the different sulphur compositions observed. We also report that differences in abundances for a given species between the static and dynamic models can reach six orders of magnitude in the hot core, which reveals the key role of the choice of model in astrochemical studies.
\end{abstract}



\section{Introduction}

Hot cores have first been defined as small ($< 0.1$ pc), dense ($n_H > 2\times10^7$ cm$^{-3}$) and warm ($T>100$ K) regions surrounding high-mass protostars in their early phase of formation \citep[see for example][]{Kurtz00,VDTak04}. Because of their high temperatures, these regions are characterized by the sublimation of icy mantles of dust grain. Therefore they present high abundances of hydrogenated molecules such as water (H$_2$O), hydrogen sulfide (H$_2$S), or complex organic molecules such as methanol \citep[CH$_3$OH, see][and references therein]{Schoier02}. These molecules are originally synthesized on dust grains in the cold dense cloud from which the protostar has formed. Once evaporated in the hot core, they undergo further gas-phase chemical reactions \citep{Wakelam04,Garrod06,Herbst09}. It is now generally admitted that low-mass protostars present the same kind of physico-chemical structure called 'hot corinos' \citep{Ceccarelli96,Ivezic97}. They differ from their high-mass counterparts mainly in size and consequently in infall timescale, which could maybe impact the chemical composition. Our Sun being a low-mass star, the chemistry that takes place in these 'small' hot cores is important to understand the history of the material from which planetary systems such as ours are formed. In this paper, we use the expression 'hot core' as a generic term to designate the hot and dense regions surrounding both high-mass and low-mass protostars.  \\
To model the chemistry of hot cores, several types of simulations exist throughout the literature \citep[see for instance][]{Charnley97,Hatchell98,Garrod06,Wakelam11,Hincelin16}. Ranging from simple 0D static gas-phase models to complex 3D gas-grains ones, different assumptions are made for each type of model, regarding for instance the age of the parent cloud or its free-fall time. These numerous approaches to hot cores chemistry and the different hypotheses they imply raise the question of the uniformity of the results obtained by these models. \\

Sulphur bearing species are often used to probe the evolution of hot cores because their abundance is particularly sensitive to physical and chemical variations. For example, the ratios SO$_2$/SO, SO$_2$/H$_2$S and OCS/H$_2$S have been proposed as chemical clocks in these regions \citep{Charnley97,Hatchell98,Wakelam11}, SO is often used to trace small-scale heating process such as shocked regions \citep{Viti01,Podio15}, or the centrifugal barrier \citep{Sakai14}, and OCS can efficiently trace the infalling-rotating envelope \citep{Oya16}.\\
Hence, the modeling of sulphur chemistry in such hot and dense regions is crucial for a better understanding of the star formation process. Moreover, such models can bring new hints on the main form of sulphur in dense clouds. Indeed the gas phase abundance of atomic sulphur in the diffuse medium is observed to be constant with cloud density, around its cosmic value of 10$^{-5}$ \citep[see for instance][]{Jenkins09}. However, in dense clouds the total abundance of detected S-bearing molecules only accounts for 0.1\% of the cosmic abundance of atomic sulphur \citep{Tieftrunk94,Charnley97}. Furthermore, for most chemical models, assuming an initial abundance of sulphur as high as its cosmic value results in predicted abundances of observable S-bearing species much higher than the observed ones. Consequently, modelers usually assume a depleted abundance of sulphur compared to its cosmic value. Therefore, the main reservoirs of sulphur in dense clouds, and consequently in hot cores, are still uncertain.\\
Let's add to these uncertainties the fact that observations of S-bearing species in hot cores is also a puzzling issue, since a large variety of sulphur compositions have been observed towards different hot cores and therefore no global trend has yet been found \citep[see figure 5 of][and references therein]{Woods15}. However, a given set of hot cores can present similar sulphur composition \citep[see for example][]{Minh16}, which would suggest similar evolutionary stages.\\
Recent studies have put forward evidences in favor of the long lasting idea that because hydrogenation is the most effective chemical process in icy grain mantles, H$_2$S could be the sought main reservoir of sulphur in dense clouds \citep[see][]{Minh90,Charnley97}. \citet{Holdship16} studied the properties of H$_2$S in the low-mass protostar L1157-B1 and showed that a significant fraction of the sulphur is likely to be locked into the form of H$_2$S prior to evaporation of the grain mantles in the hot corino. Moreover, we reported the first chemical model able to reproduce observed abundances of S-bearing species in dense clouds using as initial abundance of sulphur its cosmic one \citep[or three times depleted, see][]{Vidal17} . In this study we showed that the main form of sulphur in dense clouds critically depends on the age of the cloud. We proposed that this reservoir could be atomic sulphur for clouds of ages $< 5\times10^5$ years, or a shared reservoir between HS and H$_2$S in the grain mantles for older clouds.\\	 

In this paper we aim to present a comprehensive study of the modeling of S-bearing species in hot cores. More particularly, we try to determine how the history (i.e. the pre-collapsing chemical composition) of a hot core impacts on sulphur chemistry, and how important is the choice of the type of model used for hot cores studies. Hence, we study the evolution of the abundances of the main S-bearing species observed in hot core as given by different types of simulations using several physico-chemical parameters. The gas grain model as well as the different types of simulations we run are presented in section 2. In section 3 we present the results of our hot cores simulations for sulphur chemistry, and we discuss these results regarding the issues we want to address in the last section.

\section{Model description}

In order to conduct a comprehensive study of the sulphur chemistry in hot cores, as well as highlight the discrepancies between the different models commonly used in the literature, we will use three types of simulations: 

\begin{enumerate}
\item 0D simulations for which we consider constant physical parameters throughout the simulation time,\\
\item 1D static simulations for which each cell of the spatial grid evolves with constant physical parameter as a 0D model,\\
\item 0D dynamic simulations for which each cell of the spatial grid evolves with time-dependant physical parameters.
\end{enumerate}

For all the simulations, we use the latest version of the time dependent gas-grain \textsc{Nautilus v-1.1} chemical model \citep{Ruaud16}. In the following, we describe the chemical model, then give the parameters of the parent cloud model we use to get the same pre-collapsing chemical composition for all hot cores models. We describe afterwards the different parameters we use for each simulation.

\subsection{Chemical Model} \label{chemmod}

The \textsc{Nautilus} chemical model computes the evolution of chemical abundances for a given set a physical and chemical parameters. It can simulate a three-phase chemistry including gas-phase, grain-surface and grain bulk chemistry, along with the possible exchanges between the different phases \citep{Ruaud16}. These exchanges are:  the adsorption of gas-phase species onto grain surfaces, the thermal and non-thermal desorption of species from the grain surface into the gas-phase, and the surface-bulk and bulk-surface exchange of species. The chemical desorption process used in the model is the one depicted in \citet{Garrod07}. They considered that for each surface reaction leading to a single product, a part of the energy released by the reaction can contribute to the desorption of the product in the gas-phase using the Rice-Ramsperger-Kessel theory. The fraction of the product desorbed in the gas-phase depends on the binding energy of the product to the surface, the enthalpy of the reaction, and the fraction of the released energy that is lost to the surface. In our case, we use a $a$ parameter of 0.001, which produces approximately a 1\% efficiency evaporation for all species. Moreover, the grain chemistry takes into account the standard direct photodissociation by photons along with the photodissociation induced by secondary UV photons introduced by \citet{Prasad83}. These processes are effective on the surface as well as in the bulk of the grains. The model also takes into account the newly implemented competition between reaction, diffusion and evaporation as suggested by \citet{Chang07} and \citet{Garrod11}. The diffusion energies of each species are computed as a fraction of their binding energies. We take for the surface a value of this ratio of 0.4 as suggested by experiments and theoretical work made on H \citep[see][and reference therein]{Ruaud16}, CO and CO$_2$ \citep[see][]{Karssemeijer14}. This value is then extrapolated to every species on the surface. For the bulk we take a value of 0.8 \citep[see also][]{Ruaud16}.\\

Given the high temperature regimes encountered in this study, we use the \textit{ad hoc} formation mechanism for H$_2$ described in \citet{Harada10}. They consider that the formation rate of H$_2$ can be written as:

\begin{eqnarray}
	\frac{dn(\text{H$_2$})}{dt}=\frac{1}{2}n_Hv_Hn_g\sigma_gS(T)\epsilon
	\label{eq_0}
\end{eqnarray}\\

where $n_H$ is the number density and $v_H$ the thermal velocity of hydrogen atoms, respectively in part.cm$^{-3}$ and cm.s$^{-1}$, $n_g$ is the number density of grains in part.cm$^{-3}$, $\sigma_g$ is the cross section of a grain in cm$^{2}$, S is the sticking coefficient for hydrogen atom as a function of temperature and $\epsilon$ is the recombination efficiency. For our study, we use for the recombination efficiency the results from \citet{Cazaux05}, and for the sticking coefficient the expression derived by \citet{Chaabouni12}.\\

The reference chemical network is \textit{kida.uva.2014} \citep[see][]{Wakelam15} for the gas-phase and the one described in \citet{Ruaud16} for the grains. To this was added the sulphur network detailed in \citet{Vidal17} \citep[including the reactions given in][]{Druard12}, as well as the chemical schemes for carbon chains proposed in \citet{Wakelam15b}, \citet{Loison16}, \citet{Hickson16} and \citet{Loison17}. Note that all abundances in this paper are expressed with respect to the total H density.

\subsection{The parent cloud parameters} 

In order to model the chemistry of a given hot core, one must consider as initial condition the chemical composition of its parent cloud before it collapses. Moreover, if we want to be able to compare the outputs of our different types of simulations, it makes sense to use the same initial chemical composition. Hence, we begin to run a simulation with the commonly used dark clouds physical parameters, namely a gas and dust temperature of 10 K, an atomic hydrogen total density of $2\times10^4$ cm$^{-3}$, a cosmic ionization rate of $1.3\times10^{-17}$~s$^{-1}$, and a visual extinction of 15 mag. Our set of initial abundances is summarized in table \ref{tab_1}. We start with all species in their atomic (or ionized) form, except for hydrogen which is assumed to be entirely in its molecular form. As our model does not require additional depletion of sulphur from its cosmic value to reproduce dark clouds observations \citep{Vidal17}, we choose to use it as initial abundance of sulphur.\\

\begin{table}
\caption{Initial abundances. *$a(b)$ stands for $a\times10^b$.}
	\begin{center}
		\begin{tabular}{l r r}
		\hline
		\hline
   		Element & $n_i/n_H$* & References \\
   		\hline
		H$_2$    & 0.5             &    \\
   		He          & 0.09           & 1 \\
		N            & 6.2(-5)       &  2 \\
		O            & 2.4(-4)       &  3 \\
		C$^+$    & 1.7(-4)       &  2 \\
		S$^+$    & 1.5(-5)       &  2 \\
		Si$^+$   & 8.0(-9)       &  4 \\
		Fe$^+$  & 3.0(-9)       & 4\\
		Na$^+$  & 2.0(-9)       & 4 \\
		Mg$^+$ & 7.0(-9)       & 4\\
		P$^+$   & 2.0(-10)     & 4\\
		Cl$^+$  & 1.0(-9)       & 4 \\
		F           & 6.7(-9)     & 5 \\
		\hline
 		\end{tabular}
	\end{center}
	\medskip{(1) \citet{Wakelam08}, (2) \citet{Jenkins09}, (3) \citet{Hincelin11}, (4) Low-metal abundances from \citet{Graedel82}, (5) Depleted value from \citet{Neufeld05}}
  	\label{tab_1}
\end{table}

We showed that the evolution time of the parent cloud is critical for the sulphur reservoirs \citep{Vidal17} and the same kind of results is observed for oxygen in our simulations. Hence, in order to study the importance of the chemical history of the parent cloud on the hot cores composition, we use the outputs of the parent cloud simulation at two different final times for all our simulations: $10^5$ and $10^6$ years. Both these ages are acceptable for dark clouds and allow us to get two very different chemical compositions of the cloud before it collapses. On the one hand, in the case of the less evolved dark cloud (hereafter noted LEDC), most of the oxygen and sulphur are still in the gas phase in atomic form (respectively 42 and 61\% of their total amount), or in the form of CO (26\%) and CS (15\%), respectively. The remainder is, for both species, locked in icy grain bulks, mainly in the form of H$_2$O, HS and H$_2$S. On the other hand, in the case of the evolved dark cloud (hereafter noted EDC), more than 95\% of the oxygen is locked in the ices mainly in the form of H$_2$O (53\%) and H$_2$CO (9\%). As for sulphur, more than 90\% is locked in the ices, including its main reservoirs HS (35\%) and H$_2$S (26\%). The two pre-collapse cloud compositions are summarized in table \ref{tab_2}. For more details on the time evolution of the abundances of the main S-bearing species in the parent cloud, see figure 1 of \citet{Vidal17}. 

\begin{table}
\caption{Description of the pre-collapse dark cloud oxygen and sulphur composition. The prefix 'b-' is for the bulk species.}
	\begin{center}
		\begin{tabular}{c c}
		\hline
		\hline
   		\multicolumn{2}{c}{LEDC} \\
   		\hline
		Oxygen & Sulphur\\
		\hline
		O (42\%) & S (61\%)\\
		CO (26\%) & CS (15\%)\\
		b-H$_2$O (22\%) & b-HS (6\%)\\
		b-O (1\%) & b-H$_2$S (5\%)\\
		\hline
		\hline
		\multicolumn{2}{c}{EDC}\\
		\hline
		Oxygen & Sulphur\\
		\hline
		b-H$_2$O (53\%) & b-HS (35\%)\\
		b-H$_2$CO (9\%) & b-H$_2$S (26\%)\\
		b-CO (8\%) & b-NS (17\%)\\
		b-CH$_3$OH (7\%) & b-S (8\%)\\
		\hline
 		\end{tabular}
	\end{center}
  	\label{tab_2}
\end{table}

\subsection{The 0D simulation parameters} \label{sec_0Da}

For this study, we begin to run 0D simulations with the main purpose of getting a comprehensive look at sulphur chemistry in hot core. As \citet{Charnley97} showed the important role of temperature on sulphur chemistry in such environments, we present calculations for a typical hot core density of $2\times10^7$ cm$^{-3}$, and two temperature regimes of 100 and 300 K. Hence we get four 0D simulations with different pre-evaporative compositions (EDC and LEDC) and temperatures (100 and 300 K).

\subsection{The 1D static simulation parameters} \label{sec_1Dp}

The 1D model used in this paper follow the physical structure for the envelope of the low-mass protostar IRAS 16293-2422 from \citet{Crimier10}, which was constrained through multi-wavelength dust and molecular observations. Indeed this protostar is believed to have a hot core within the $\sim 150$ AU around its centre \citep[see][]{Schoier02}. The density and temperature radial evolutions are shown in figure \ref{fig_1} (a) and (b), respectively (solid line). For this model, the spatial limit of the hot core (T > 100 K) is located at $R_{HC}$ = 80 AU (solid grey line on figure \ref{fig_1} (b)).

\subsection{The 0D dynamic simulation parameters} \label{sec_0DD}

\begin{figure}
        \begin{center}
                \includegraphics[scale=0.14]{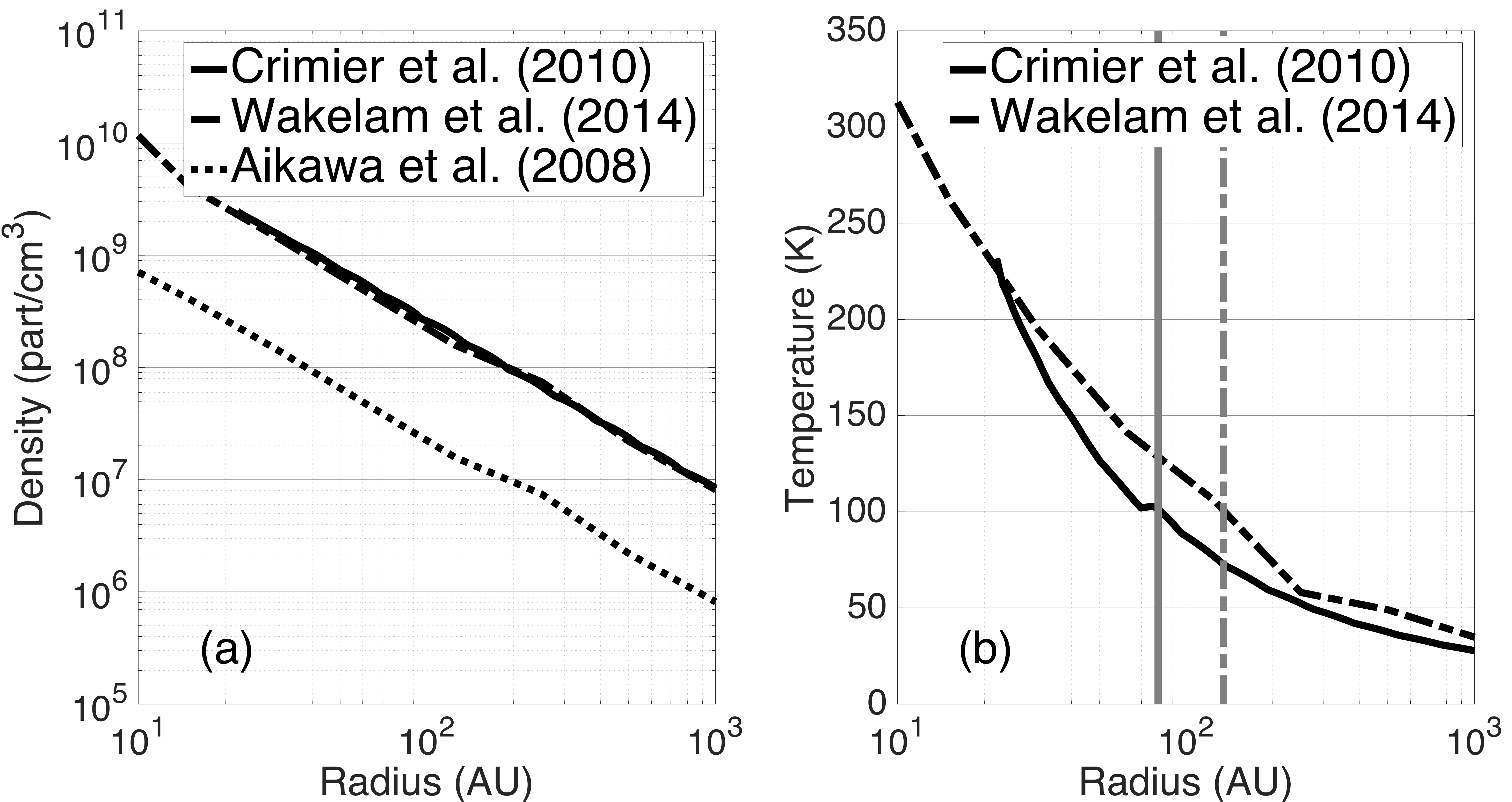}
                \caption{Radial structure of the 1D static and 0D dynamic model at final time for: (a) density, for which we plot the initial structure from \citet{Aikawa08} as reference, (b) temperature, for which we plot the limit of the hot core (T > 100 K) for the 1D static ($R_{HC}$ = 80 AU, solid grey vertical line) and 0D dynamic ($R_{HC}$ = 135 AU, dashed grey vertical line).}
                \label{fig_1}
        \end{center}
\end{figure}

The structure we use for our dynamic simulation is the same as in \citet{Aikawa08}, \citet{Wakelam14} and \citet{Majumdar16}, and was computed from the radiation hydrodynamic (RHD) model from \citet{Masunaga00}. It initially starts from a parent cloud with a central density of $\sim 6\times10^4$ cm$^{-3}$, a radius of $4\times10^4$ AU and a total mass of 3.852 M$_\odot$. Then the model follows the collapse of the prestellar core, which eventually forms a protostellar core after $2.5\times10^5$ yr. Finally, the protostar grows by mass accretion from the envelope for $9.3\times10^4$ yr. As in \citet{Wakelam14}, we have multiplied by 10 all the densities of the simulations in order for the final physical structure of the dynamic model to be similar to the 1D structure of \citet{Crimier10}. The consequence of this modification will be discussed in section \ref{sec_discdens} \citep[see also section 4.5 of][]{Wakelam14}. Figure \ref{fig_1} (a) shows the resulting final density radial evolution (dashed line) as well as the previous one (dotted line), and figure \ref{fig_1} (b) the final temperature radial evolution (dashed line). For this model, the spatial limit of the hot core (T > 100 K) is located at $R_{HC}$ = 135 AU (dashed grey line on figure \ref{fig_1} (b)).\\

The designations and physical parameters of all the simulations presented in this paper are summarized in table \ref{tab_3}.

\begin{table*}
\caption{Summary of the simulations designations and physical parameters.}
	\begin{center}
		\begin{tabular}{c c c}
		\hline
		\hline
		Simulations & Physical parameters & Pre-collapse evolution time\\
		\hline
   		\multicolumn{3}{c}{0D Static models} \\
   		\hline
		0DS100LEDC & T = 100 K, $n_H$ = 2$\times10^7$ cm$^{-3}$ & 10$^5$ yrs\\
		0DS300LEDC & T = 300 K, $n_H$ = 2$\times10^7$ cm$^{-3}$ & 10$^5$ yrs\\
		0DS100EDC & T = 100 K, $n_H$ = 2$\times10^7$ cm$^{-3}$ & 10$^6$ yrs\\
		0DS300EDC & T = 300 K, $n_H$ = 2$\times10^7$ cm$^{-3}$ & 10$^6$ yrs\\
		\hline
		\multicolumn{3}{c}{1D Static models} \\
		\hline
		1DSLEDC & Structure from \citet{Crimier10} & 10$^5$ yrs\\
		1DSEDC & Structure from \citet{Crimier10} & 10$^6$ yrs\\
		\hline
		\multicolumn{3}{c}{0D Dynamic models} \\
		\hline
		0DDLEDC & Modified structure from \citet{Aikawa08} & 10$^5$ yrs\\
		0DDEDC & Modified structure from \citet{Aikawa08} & 10$^6$ yrs\\
		\hline
 		\end{tabular}
	\end{center}
  	\label{tab_3}
\end{table*}

\section{Results}

\subsection{0D models} \label{sec_0D}

In this section we aim to do a comprehensive study of sulphur chemistry in environments such as hot cores. In particular, we want to address the importance of the temperature as well as the pre-collapse chemical composition using the four 0D simulations defined in section \ref{sec_0Da}. Sulphur chemistry in the hot gas phase is known to be intertwined with the distribution of reactive oxygen: O, O$_2$ and OH \citep{Charnley97,Wakelam04,Esplugues14}. Hence we begin with a description of the chemistry of these species, then study the chemistry of the main S-bearing species observed towards hot and dense objects: SO, SO$_2$, H$_2$S, OCS, CS and H$_2$CS.
 
\subsubsection{Oxygen chemistry}

\begin{figure*}
        \begin{center}
                \includegraphics[scale=0.13]{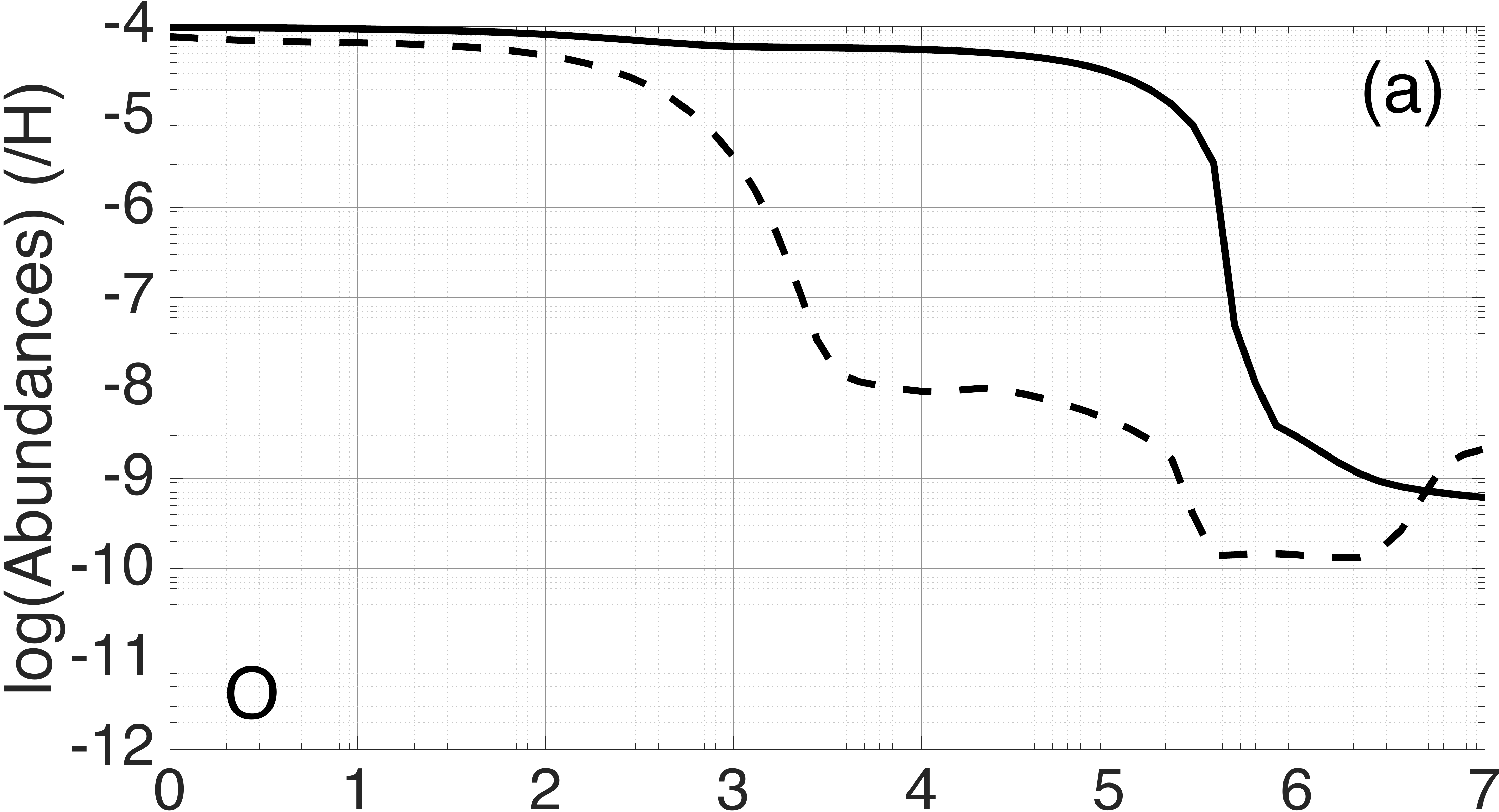}
                \includegraphics[scale=0.13]{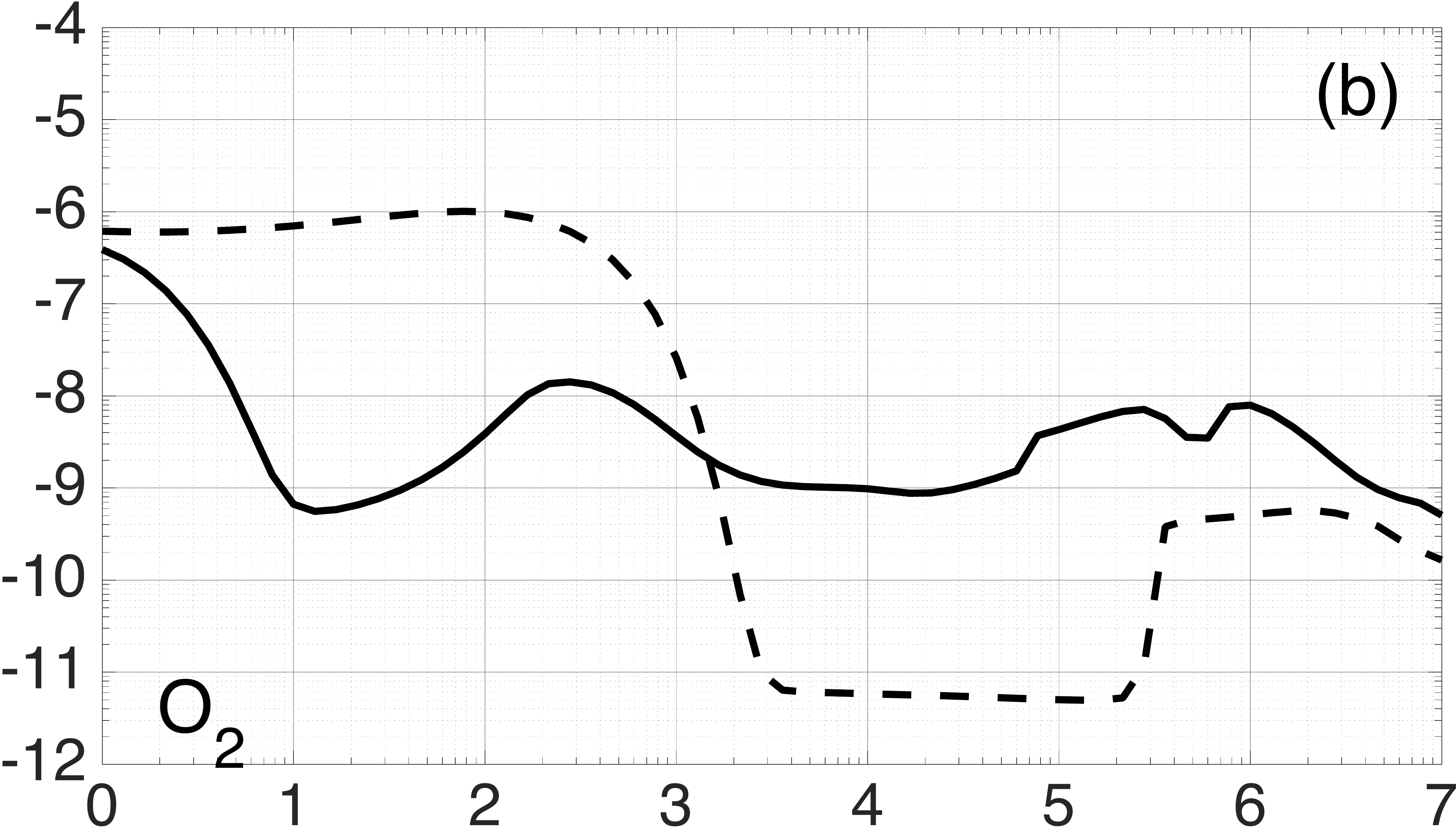}
                \includegraphics[scale=0.13]{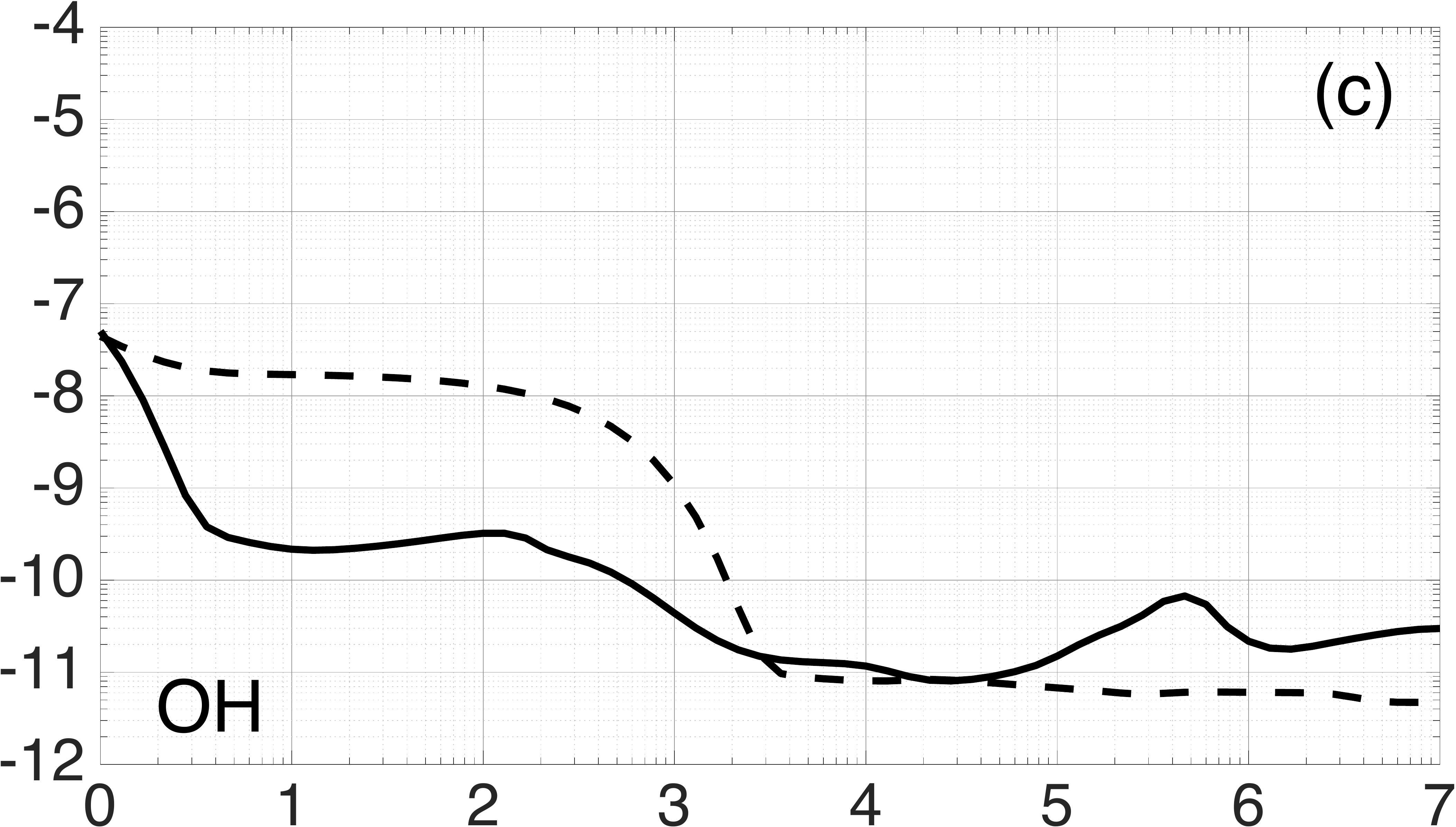}\\
                \includegraphics[scale=0.13]{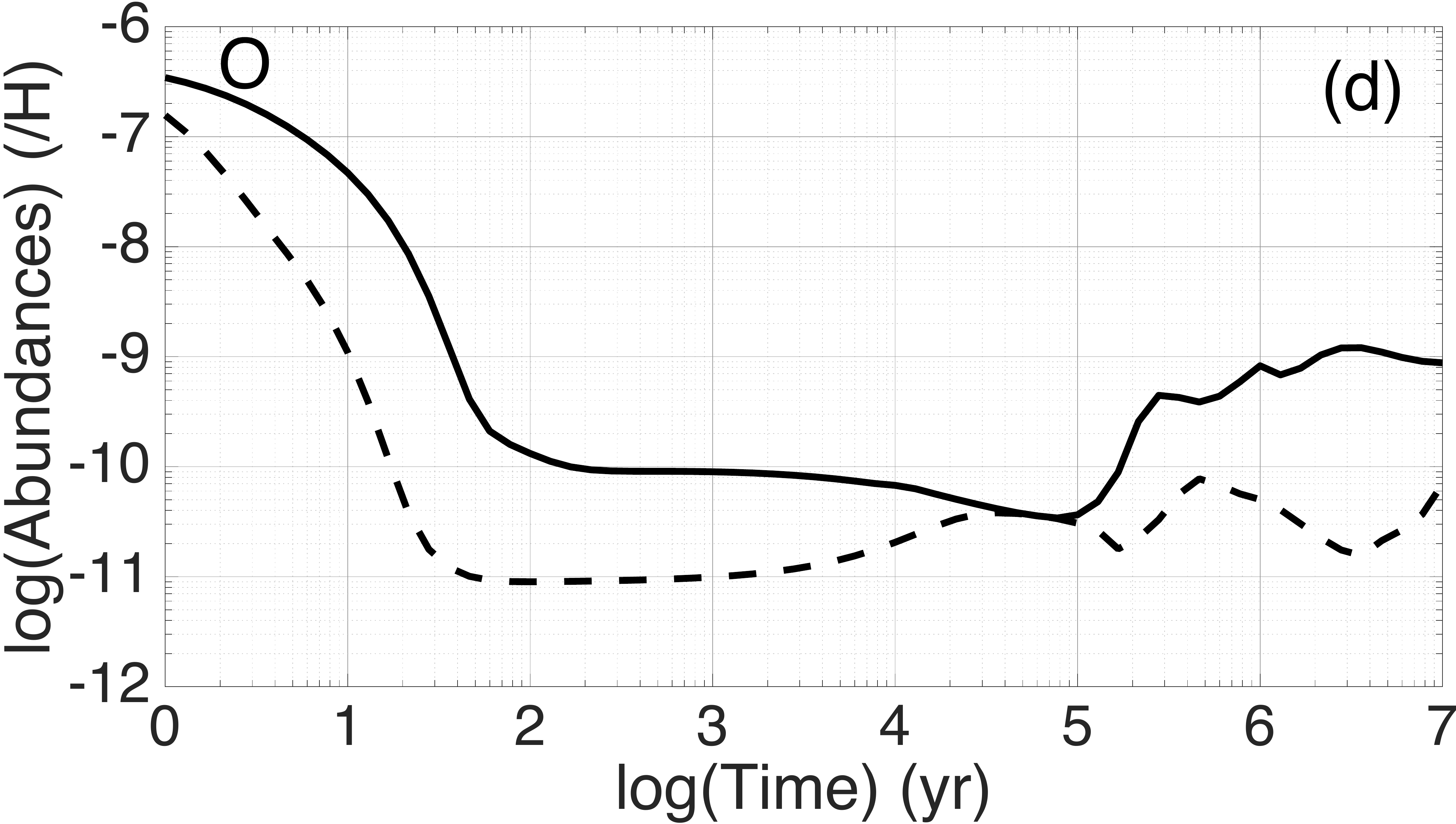}
                \includegraphics[scale=0.13]{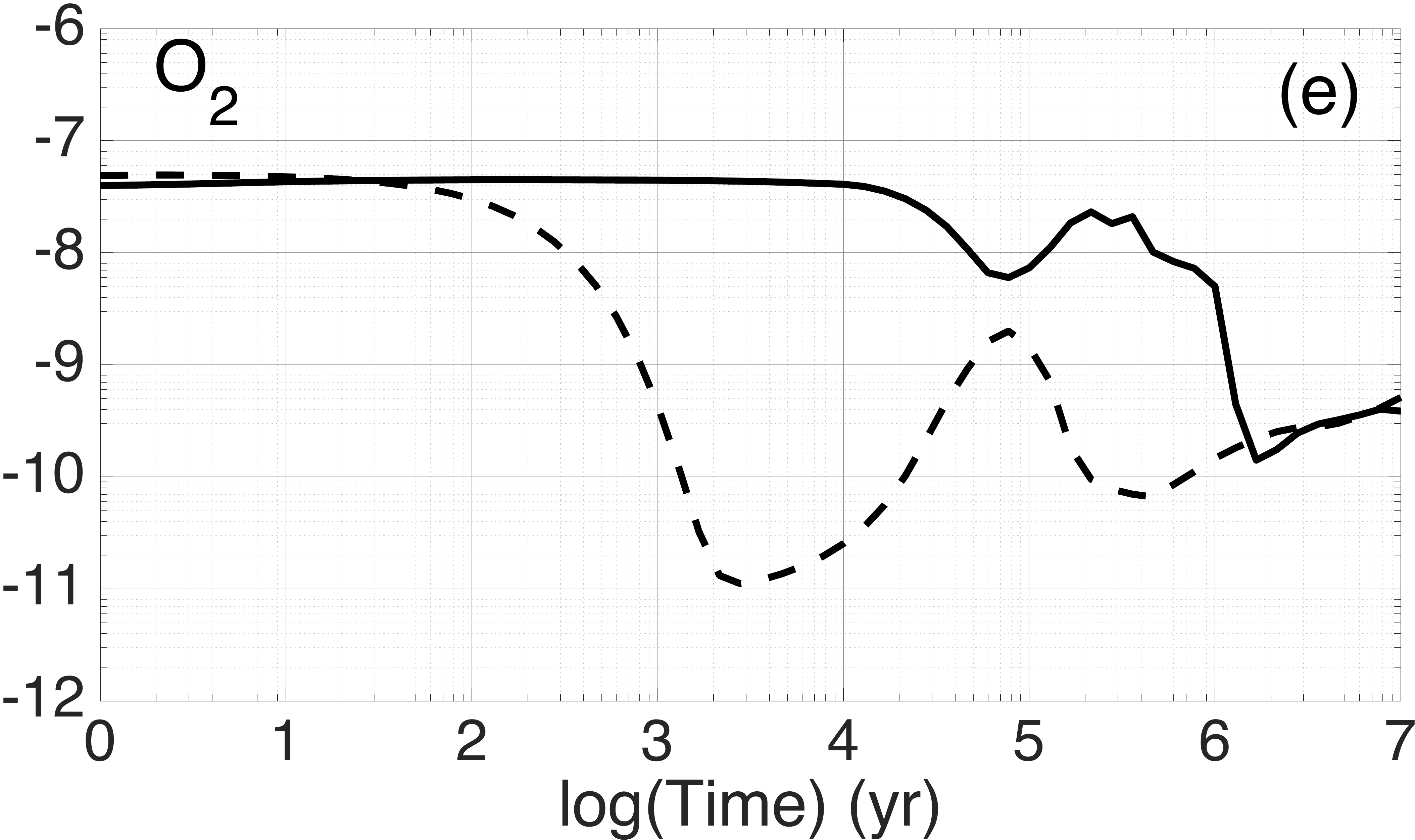}
                \includegraphics[scale=0.13]{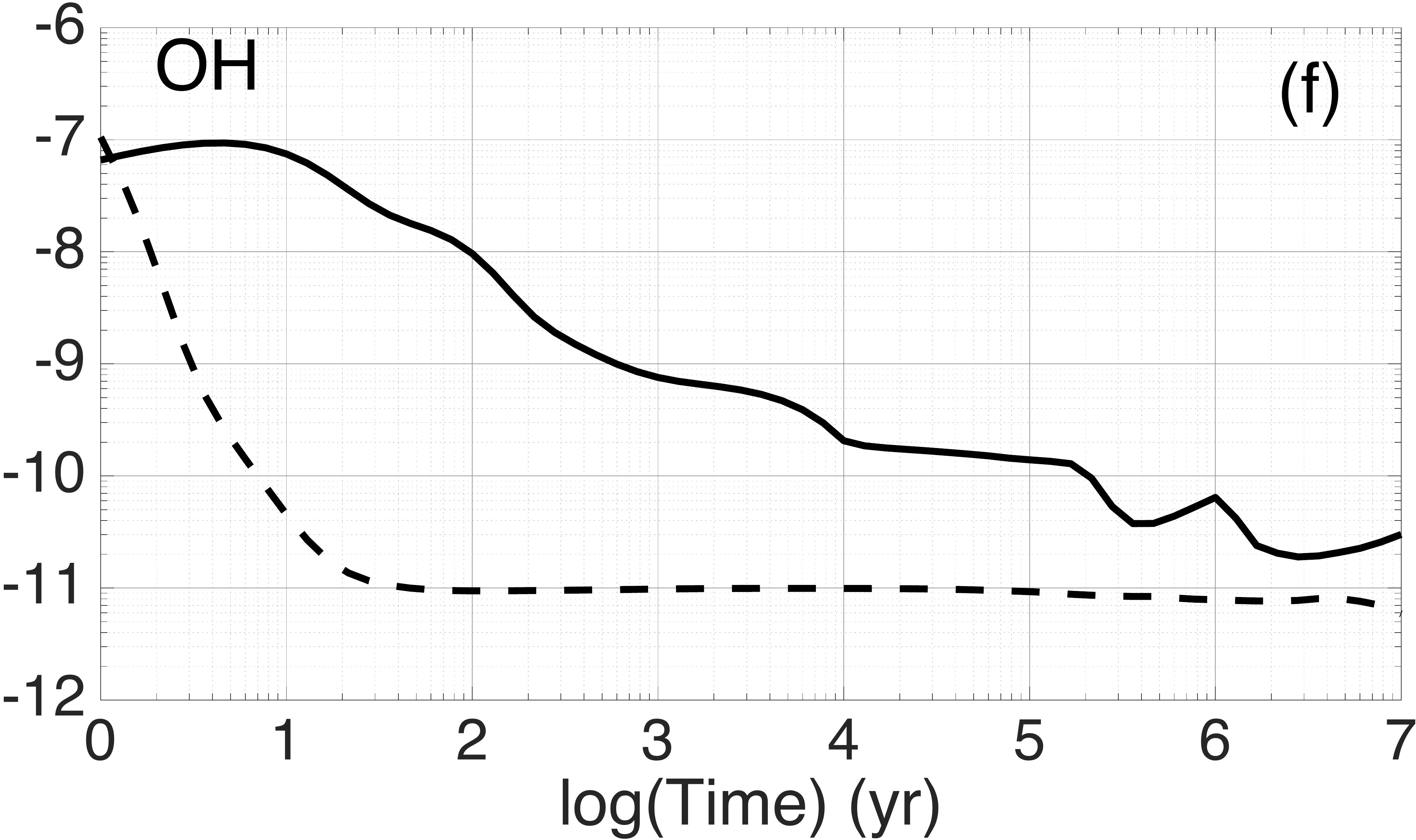}
                \caption{Abundances of O, O$_2$ and OH relative to H as a function of time for hot cores conditions: $n_H = 2\times10^7$ cm$^{-3}$ and T = 100 K (solid line) or 300 K (dashed line), and for LEDC (top panel) and EDC (bottom panel) pre-collapse compositions.}
                \label{fig_2}
        \end{center}
\end{figure*}

The left, middle and right panels of figure \ref{fig_2} show the abundances of respectively O, O$_2$ and OH, for the LEDC (top) and EDC (bottom) pre-collapse compositions. For atomic oxygen in the 0DS100LEDC case (cf figure \ref{fig_2} (a)), the temperature is not high enough for all the mantle of grain to evaporate and O reacts mainly with the S-bearing species available in the gas phase. It initially reacts with CS which is the second most abundant S-bearing species in the gas-phase, to form S and CO. As the abundance of CS consequently decreases, O is then mainly destroyed through formation of SO and SO$_2$:

\begin{eqnarray}
	\text{O} + \text{S$_2$} &\to &\text{S} + \text{SO} \label{eq_1a}\\
	\text{O} + \text{SO} &\to & \text{SO$_2$} \label{eq_1b}
\end{eqnarray}

In the 0D300LEDC case (cf figure \ref{fig_2} (a)), the temperature of the core is high enough for mantle hydrocarbons (H$_x$C$_n$) to evaporate. They participate in the consumption of O by forming mainly CO, causing the O abundance to decrease much faster than in the 0D100LEDC case. However, the most efficient reaction at this temperature is:

\begin{eqnarray}
	\text{O} + \text{H$_2$} &\to &\text{H} + \text{OH}
	\label{eq_2}
\end{eqnarray}

Towards the end ($t > 10^6$ yrs), the abundance of atomic oxygen increases again from the reaction:

\begin{eqnarray}
	\text{CO} + \text{He$^+$} &\to &\text{He} + \text{O} + \text{C+}
	\label{eq_3}
\end{eqnarray}

as well as photodissociation by secondary UV photons of SO and SO$_2$.\\
With the pre-collapse composition of the EDC, contrary to the LEDC, most of the oxygen is trapped in grain mantle, mainly under the form of H$_2$O, H$_2$CO and CO (see table \ref{tab_2}). Hence, few atomic oxygen is available in the gas phase even after sublimation of the grain mantle, which explains that its initial abundance in both EDC cases is lower by two orders of magnitude compared to LEDC ones. In the 0DS100EDC case (cf figure \ref{fig_2} (d)), the temperature of the corino is hot enough to sublimate light species, especially HS and NS which are two of the sulphur reservoirs in the pre-collapse composition. These species react rapidly with O to form SO and NO for $10^4$ years. After this time, it is HCNS which reacts mainly with atomic oxygen through the following reactions:

 \begin{eqnarray}
	\text{O} + \text{HCNS} &\to &\text{HCN} + \text{SO} \label{eq_4a}\\
	&\to & \text{HCO} +  \text{NS}
	\label{eq_4b}
\end{eqnarray}

The global increase of abundance of O at the end of the simulation is due to reaction (\ref{eq_3}).\\

In the 0D300EDC case (cf figure \ref{fig_2} (d)), most of the sublimated hydrocarbons efficiently form CH$_3$ which consumes O in the first hundred years following:

 \begin{eqnarray}
	\text{O} + \text{CH$_3$} &\to &\text{H} + \text{CO} + \text{H$_2$}\\
	&\to & \text{H} +  \text{H$_2$CO}
	\label{eq_5}
\end{eqnarray}

Afterwards, O abundance increases slowly, mainly from the reaction:

\begin{eqnarray}
	\text{NH} + \text{NO} &\to &\text{H} + \text{O} + \text{N$_2$}
	\label{eq_6}
\end{eqnarray}\\

In both 0DS100LEDC and 0DS300LEDC models, dioxygen chemistry is strongly linked with S and OH chemistry via:

\begin{eqnarray}
	\text{S} + \text{O$_2$} &\to &\text{O} + \text{SO} \label{eq_7a}\\
	\text{O} + \text{OH} &\to & \text{H} + \text{O$_2$} \label{eq_7b}
\end{eqnarray}

At 100 K (cf figure \ref{fig_2} (b)), atomic carbon initially destroys O$_2$ faster than it is created by reaction (\ref{eq_7b}) through:

 \begin{eqnarray}
	\text{C} + \text{O$_2$} &\to & \text{O} + \text{CO}
	\label{eq_8}
\end{eqnarray}

Then, atomic sulphur consumes both O$_2$ and OH (cf figure \ref{fig_2} (c)) within a timescale of a thousand years, respectively via reaction (\ref{eq_7a}) and:

 \begin{eqnarray}
	\text{S} + \text{OH} &\to & \text{H} + \text{SO}
	\label{eq_9}
\end{eqnarray}

Afterwards, both species chemistries are linked by reaction (\ref{eq_7b}). The short increase of both abundances near 10$^6$ years is mainly due to the formation of OH via electronic recombination of HOCS$^+$ and HSO$_2^+$.\\
At 300 K however (cf figure \ref{fig_2} (b)), atomic carbon is initially rapidly consumed by the two evaporated hydrocarbons C$_4$H$_2$ and C$_2$H$_2$, and does not destroys O$_2$ as efficiently as at 100 K. Instead, OH is formed rapidly via reaction (\ref{eq_2}) (cf figure \ref{fig_2} (c)), which causes the O$_2$ abundance to increase through reaction (\ref{eq_7b}) in the first hundred years. Then O$_2$ and OH are consumed in a few thousands years respectively by reaction (\ref{eq_7a}) and:

 \begin{eqnarray}
	\text{H$_2$} + \text{OH} &\to & \text{H} + \text{H$_2$O}
	\label{eq_10}
\end{eqnarray}

It should be noted that the increase in O$_2$ abundance near 10$^5$ years is due to the following ion-neutral reactions:

\begin{eqnarray}
	\text{S} + \text{O$_2^+$} &\to & \text{O$_2$} + \text{S$^+$}\\
	\text{SO$_2$} + \text{He$^+$} &\to &\text{He} + \text{O$_2$} + \text{S$^+$}
	\label{eq_11}
\end{eqnarray}\\

In the EDC cases, atomic oxygen is not abundant enough in the gas phase for reaction (\ref{eq_7b}) to be efficient. Hence, there is no evident link between O$_2$ and OH chemistries as in the LEDC cases. For the 0D100EDC model (cf figure \ref{fig_2} (e)), the abundance of dioxygen first slowly grow for a few hundred years from the reaction:

 \begin{eqnarray}
	\text{HCO} + \text{O$_2$H} &\to & \text{O$_2$} + \text{H$_2$CO}
	\label{eq_12}
\end{eqnarray}

O$_2$H is also destroyed by H to form OH, and when there is not enough left in the gas phase, HCO then reacts with O$_2$ causing its abundance to decrease after a few thousands years. Moreover, reaction (\ref{eq_7a}) also becomes efficient after 10$^4$ years and O$_2$ abundance starts to drop at this time. The increase near 10$^5$ years is manly due to reaction (\ref{eq_11}).\\
In the 0D300EDC (cf figure \ref{fig_2} (e)), the CH$_3$ formed from the evaporated hydrocarbon destroys efficiently O$_2$ for a few thousands years via:

 \begin{eqnarray}
	\text{O$_2$} + \text{CH$_3$} &\to &\text{H$_2$O} + \text{HCO} 
	\label{eq_13}
\end{eqnarray}

Afterwards, the abundance of O$_2$ globally increases from the following reactions:

 \begin{eqnarray}
 	\text{O$_3$} + \text{S} &\to &\text{SO} + \text{O$_2$} \\
	\text{O$_2$H} + \text{CH$_3$} &\to &\text{CH$_4$} + \text{O$_2$} 
	\label{eq_14}
\end{eqnarray}

OH abundance generally decreases in both EDC models. For the 0D100EDC case (cf figure \ref{fig_2} (f)), this decrease is mainly due to:

 \begin{eqnarray}
 	\text{CS} + \text{OH} &\to &\text{H} + \text{OCS} \label{eq_15a}\\
	\text{H$_2$CO} + \text{OH} &\to &\text{H$_2$O} + \text{HCO} 
	\label{eq_15b}
\end{eqnarray}

As for the 0D300EDC (cf figure \ref{fig_2} (f)), it is mainly due to reaction (\ref{eq_10}) and (\ref{eq_15b}).

\subsubsection{Sulphur chemistry}

In the following, we take a comprehensive look at the chemistry of the main neutral S-bearing species detected in hot cores namely SO, SO$_2$, OCS, H$_2$S, H$_2$CS and CS.\\~\\

\begin{figure*}
        \begin{center}
                \includegraphics[scale=0.13]{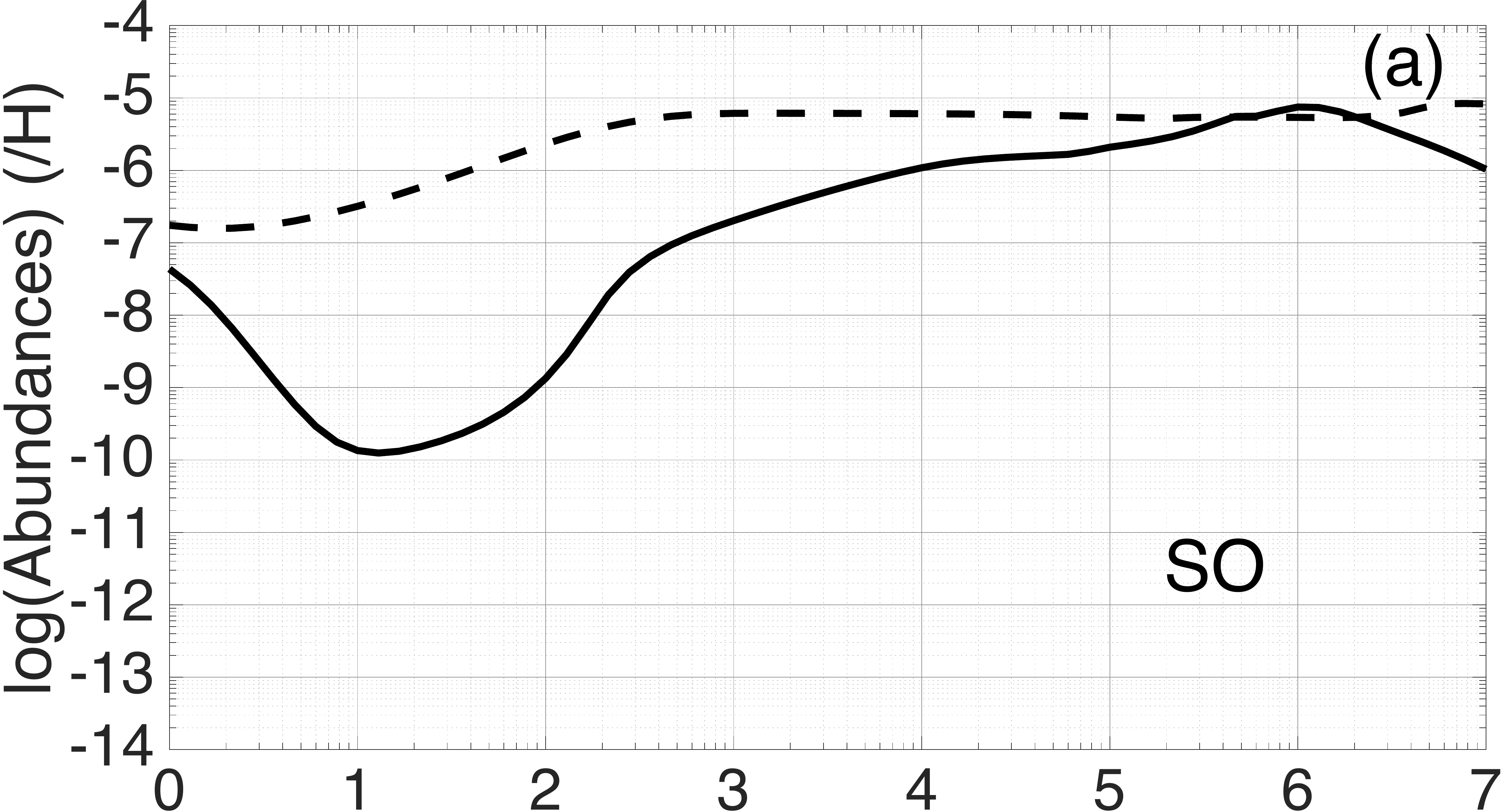}
                \includegraphics[scale=0.13]{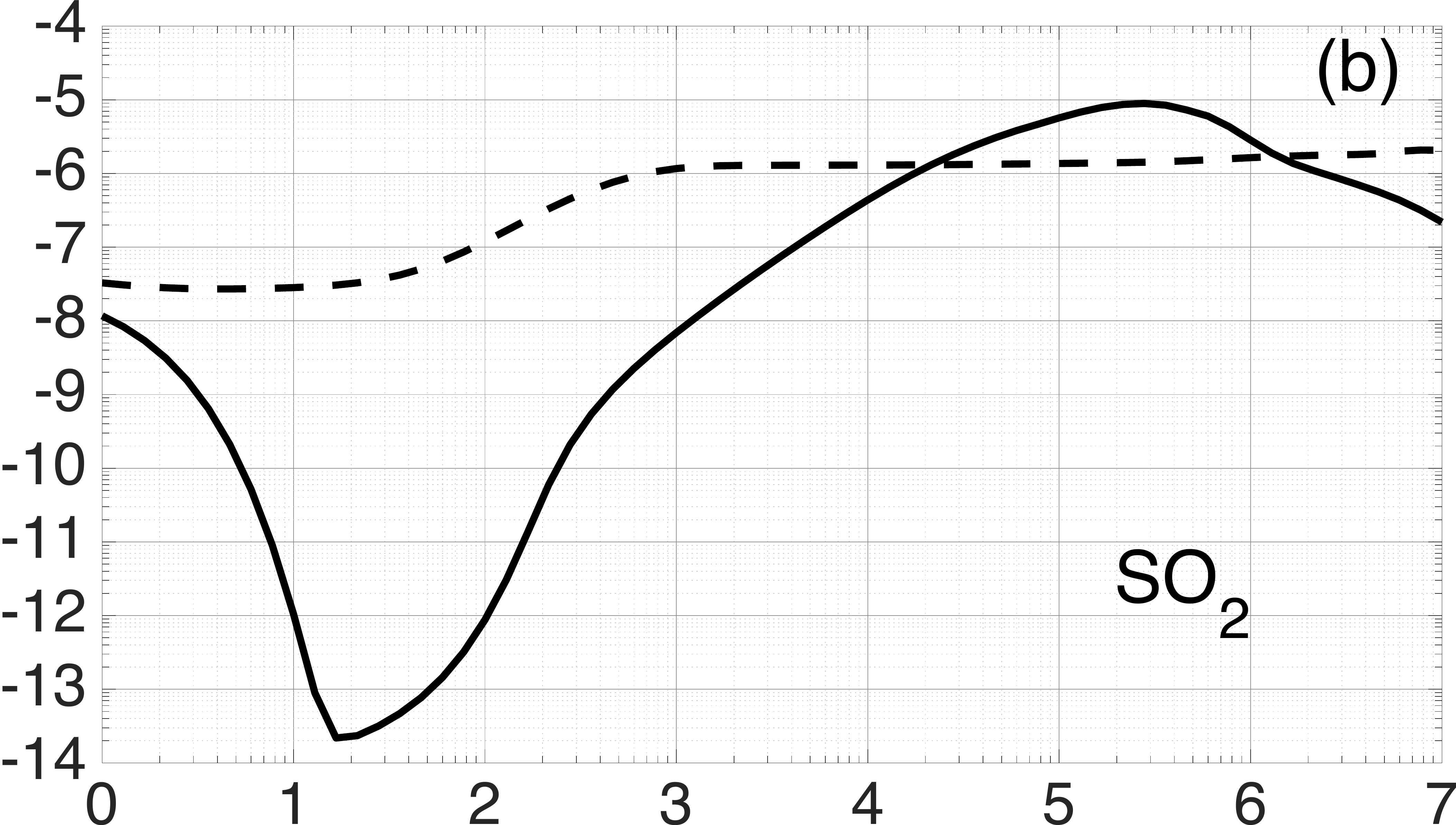}
                \includegraphics[scale=0.13]{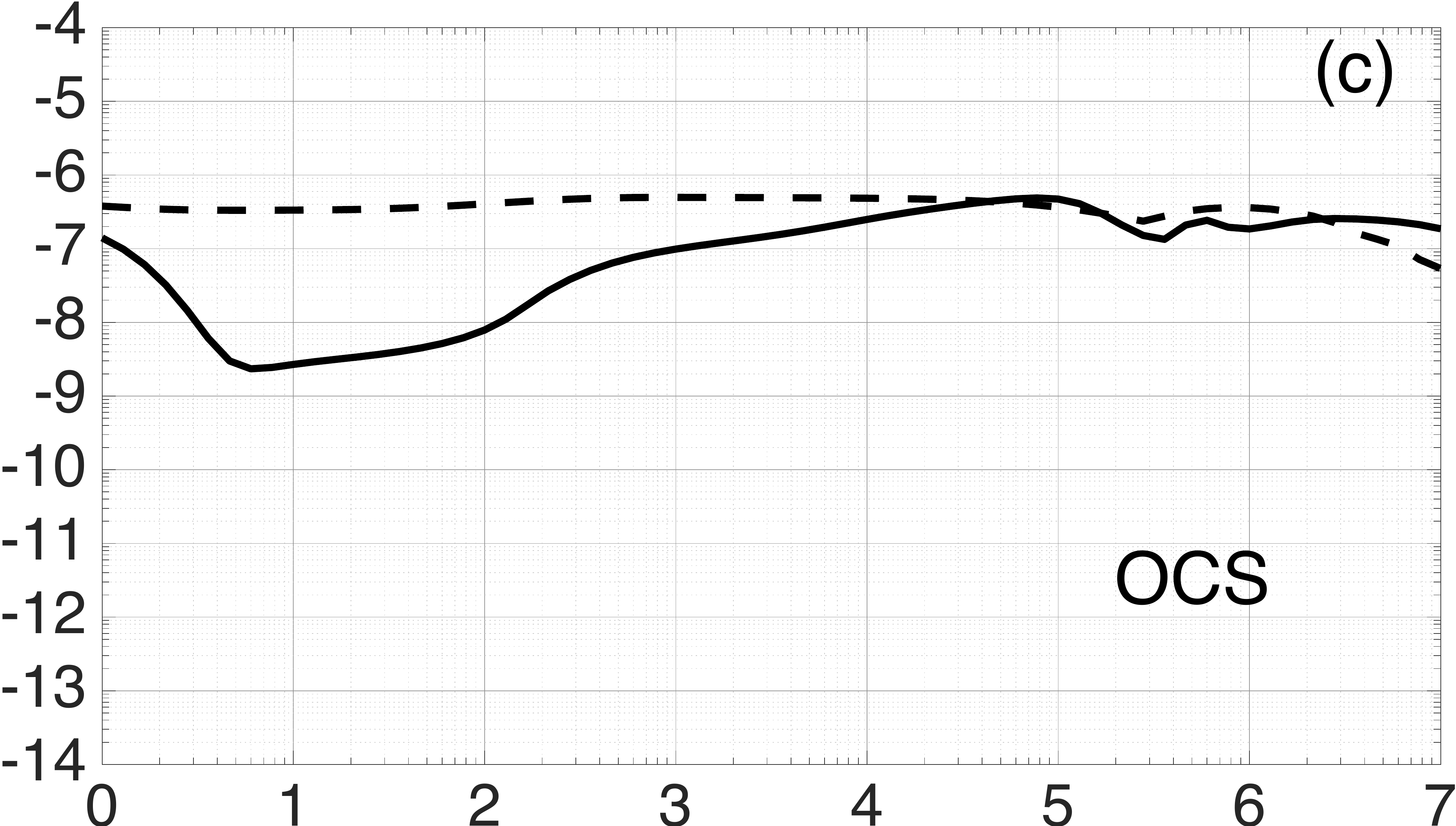}\\
                \includegraphics[scale=0.135]{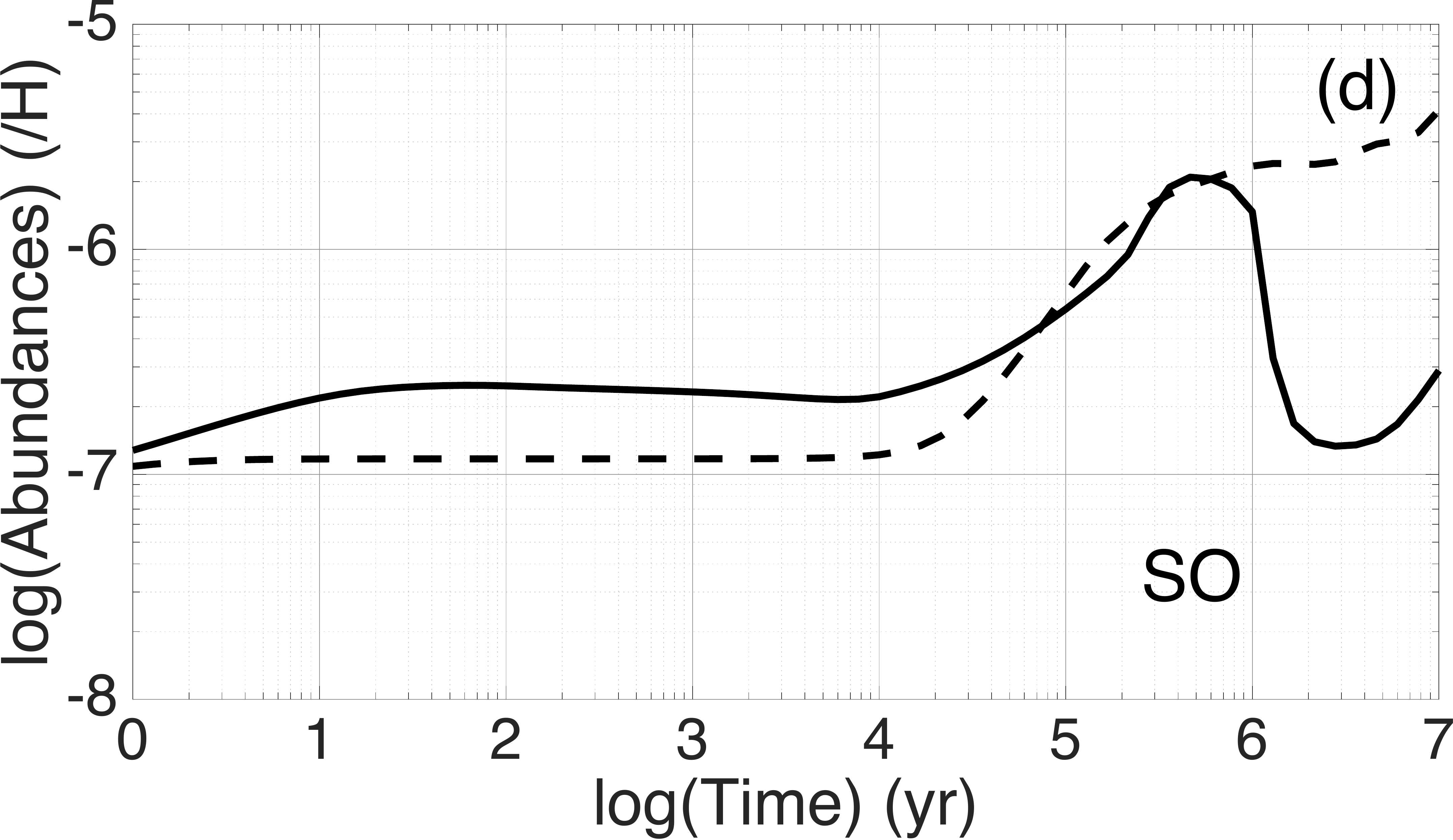}
                \includegraphics[scale=0.135]{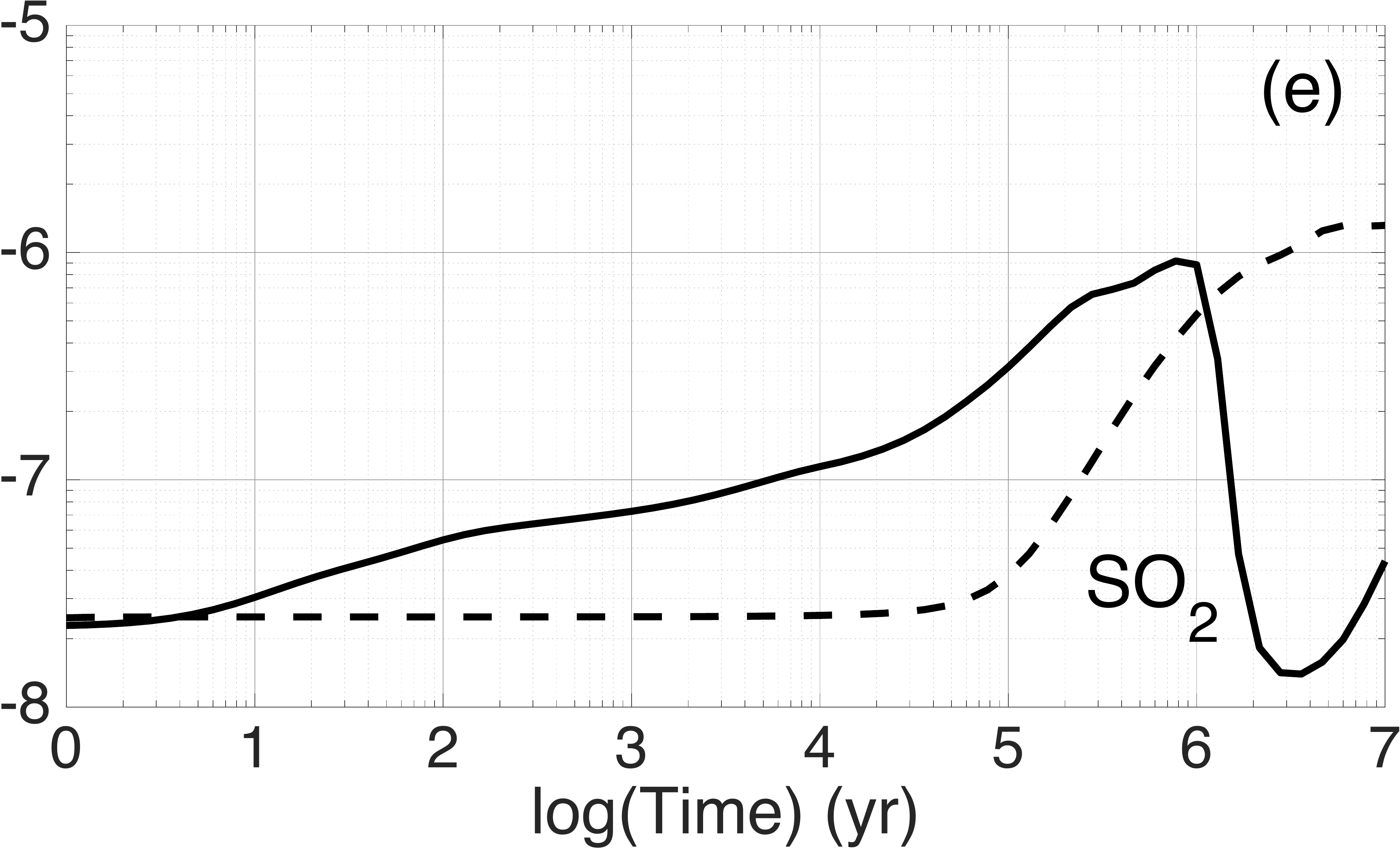}
                \includegraphics[scale=0.135]{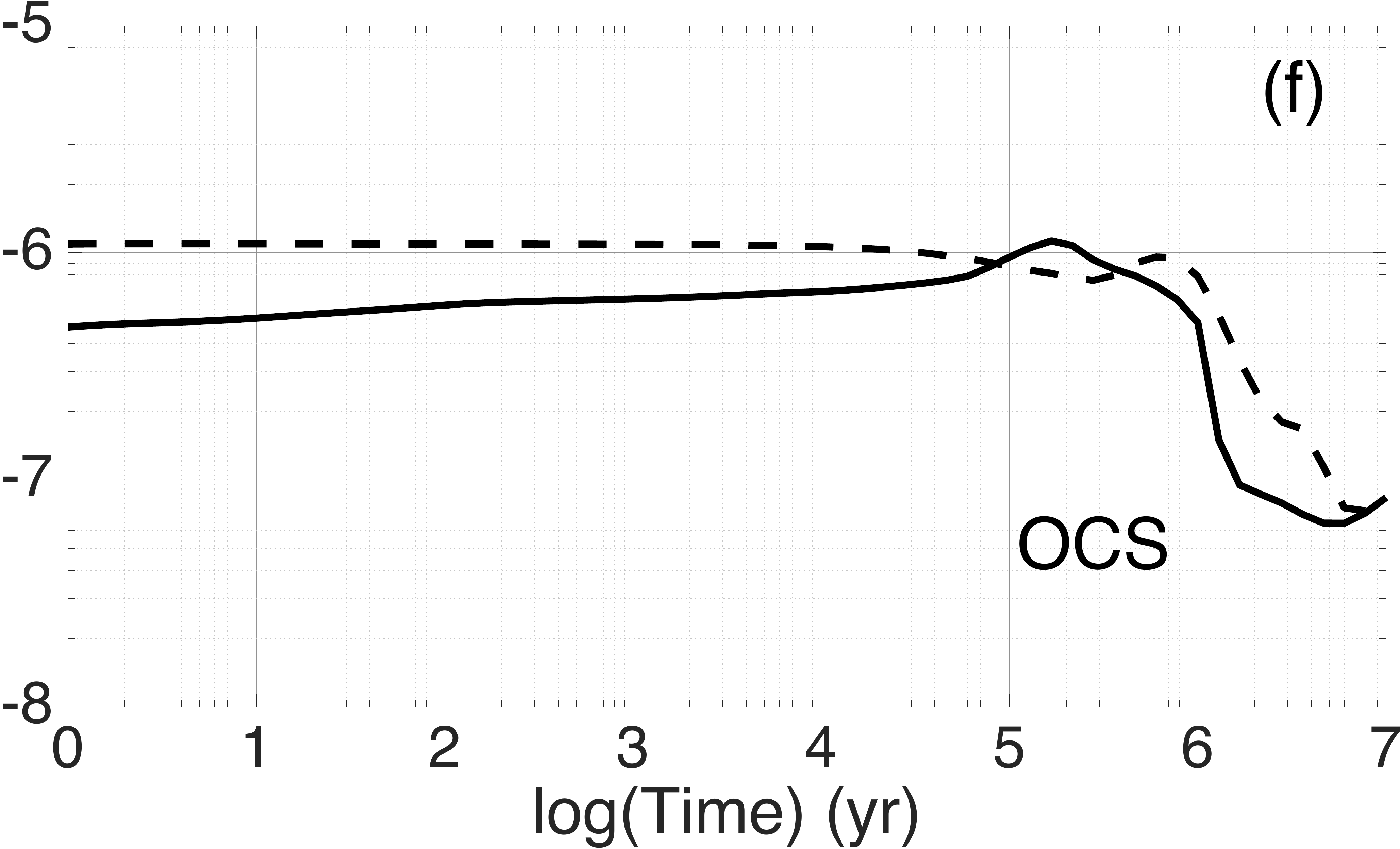}
                \caption{Abundances of SO, SO$_2$ and OCS relative to H as a function of time for hot cores conditions: $n_H = 2\times10^7$ cm$^{-3}$ and T = 100 K (solid line) or 300 K (dashed line), and for LEDC (top panel) and EDC (bottom panel) pre-collapse compositions.}
                \label{fig_3}
        \end{center}
\end{figure*}

3.1.2.1 \quad SO, SO$_2$ and OCS in the LEDC cases\\

Figure \ref{fig_3} displays the abundances of SO, SO$_2$ et OCS in both LEDC (top panels) and EDC (bottom panels) cases. In the 0DS100LEDC case, these three species are initially destroyed by atomic carbon explaining their respective drops in the first 10 years of the simulation:

 \begin{eqnarray}
 	\text{SO} + \text{C} &\to &\text{S} + \text{CO} \label{eq_16a}\\
	&\to &\text{O} + \text{CS}\label{eq_16b}\\ 
	\text{SO$_2$} + \text{C} &\to &\text{SO} + \text{CO} \label{eq_16c}\\
	\text{OCS} + \text{C} &\to &\text{CS} + \text{CO} 
	\label{eq_16d}
\end{eqnarray}

SO abundance then grows first mainly from reaction (\ref{eq_9}), then reactions (\ref{eq_1a}) and (\ref{eq_7a}) (cf figure \ref{fig_3} (a)). After 10$^6$ years, as only a small mount of reactive oxygen remains in the gas phase, SO is no longer efficiently produced. It is instead mainly destroyed by CH through:

 \begin{eqnarray}
 	\text{SO} + \text{CH} &\to &\text{H} + \text{OCS} \label{eq_17a}\\
	 &\to &\text{CO} + \text{HS} \label{eq_17b}
\end{eqnarray}

SO$_2$ is linked with SO mainly by reaction (\ref{eq_1b}), as well as:

 \begin{eqnarray}
 	\text{SO} + \text{OH} &\to &\text{H} + \text{SO$_2$} 
	\label{eq_18}
\end{eqnarray}

which is only efficient when OH is abundant enough in the gas phase (i.e. at the early beginning of the simulation, cf figure \ref{fig_2} (c)). Hence, SO$_2$ abundance grows from 10 to a few 10$^5$ years from these reactions (cf figure \ref{fig_3} (b)). Afterwards, SO$_2$ is destroyed by H$_3^+$ and C$^+$ following:

 \begin{eqnarray}
 	\text{SO$_2$} + \text{H$_3^+$} &\to &\text{H$_2$} + \text{HSO$_2^+$} \label{eq_19a}\\
	\text{SO$_2$} + \text{C$^+$} &\to &\text{CO} + \text{SO$^+$} 
	\label{eq_19b}
\end{eqnarray}

After his consumption by atomic carbon, OCS is also efficiently formed until 10$^5$ years from HCO and HCS (cf figure \ref{fig_3} (c)):

 \begin{eqnarray}
 	\text{O} + \text{HCS} &\to &\text{H} + \text{OCS} \label{eq_20a}\\
	\text{S} + \text{HCO} &\to &\text{H} + \text{OCS} 
	\label{eq_20b}
\end{eqnarray}

For the remainder of the simulation, HCO$^+$ is formed efficiently from reaction of H$_3^+$ with CO, and reacts with OCS:

 \begin{eqnarray}
 	\text{OCS} + \text{HCO$^+$} &\to &\text{CO} + \text{HOCS$^+$} 
	\label{eq_21}
\end{eqnarray}

However OCS abundance remains stable at this time because of reaction (\ref{eq_17a}).\\

In the 0DS300LEDC case, C$_4$H$_2$ and C$_2$H$_2$ thermally desorb from grain mantle. Both these species react initially with atomic carbon with reaction rates higher of more than two orders of magnitude than those of reactions (\ref{eq_16a}), (\ref{eq_16b}), (\ref{eq_16c}) and (\ref{eq_16d}) preventing SO, SO$_2$ and OCS from abrupt initial consumptions. Therefore SO and SO$_2$ form rapidly from reactions (\ref{eq_7a}), (\ref{eq_9}) and (\ref{eq_18}) (cf figure \ref{fig_3} (a) and (b)). When the abundance of reactive oxygen drops around 10$^3$ years, their abundances will undergo only small variations. Indeed as the main reactions ruling SO chemistry are no longer efficient because of the lack of reactive oxygen in the gas phase, reactions that recycle SO via HSO$^+$ allow its abundances to reach a quasi-static regime until the end of the simulation. For instance, due to the high temperature, high abundance of evaporated H$_2$O and CO render efficient the two following reactions:

 \begin{eqnarray}
 	\text{H$_2$O} + \text{HSO$^+$} &\to &\text{H$_3$O$^+$} + \text{SO} \label{eq22a}\\
	\text{CO} + \text{HSO$^+$} &\to &\text{HCO$^+$} + \text{SO} 
	\label{eq_22b}
\end{eqnarray}

The SO thus formed is then put back in HSO$^+$:

\begin{eqnarray}
 	\text{SO} + \text{H$_3^+$} &\to &\text{H$_2$} + \text{HSO$^+$} 
	\label{eq_23}
\end{eqnarray}

As a result, SO$_2$ abundances shows only a small increase during the final part of the simulation, mainly due to reaction (\ref{eq_18}).\\
OCS abundance does not vary much in the 0DS300LEDC case (cf figure \ref{fig_3} (c)). When atomic oxygen is still abundant in the gas phase, OCS is mainly formed through reaction (\ref{eq_20a}). Afterwards OCS is destroyed by secondary UV photons to form S and CO, and by H$_3^+$ through:

 \begin{eqnarray}
 	\text{OCS} + \text{H$_3^+$} &\to &\text{CO} + \text{HOCS$^+$} 
	\label{eq_21}
\end{eqnarray}

HOCS$^+$ then recombines electronically to form either CS and OCS, which explains why the latter abundance decreases slowly.

3.1.2.2 \quad SO, SO$_2$ and OCS in the EDC cases\\

The bottom panel of figure \ref{fig_3} displays the abundances of SO, SO$_2$ and OCS in both EDC cases. On the one hand, it appears that the chemistry of these species does not depend as much on the temperature as in the LEDC cases. This can be explained by the poor abundance of reactive oxygen, especially atomic oxygen, in the gas phase in the EDC cases compared with the LEDC cases. Table \ref{tab_4} displays the amount of reactive oxygen (relative to that of total oxygen) in the gas phase at the first time step of each hot core simulation. It shows that in the LEDC cases at least 33\% of the total amount of oxygen is under reactive form in the gas phase, against at most 0.1\% in the EDC cases.\\

\begin{table}
\caption{Description of the reactive oxygen composition in the gas phase at the first time step of the hot cores simulations (post-collapse). The values display the percentage relative to the total abundance of oxygen.}
	\begin{center}
		\begin{tabular}{c c}
		\hline
		\hline
   		\multicolumn{2}{c}{LEDC} \\
   		\hline
		100 K & 300 K\\
		\hline
		41\% & 33\% \\
		\hline
		\hline
		\multicolumn{2}{c}{EDC}\\
		\hline
		100 K & 300 K\\
		\hline
		0.1\% & < 0.09\% \\
		\hline
 		\end{tabular}
	\end{center}
  	\label{tab_4}
\end{table}

On the other hand the chemistry of SO, SO$_2$ and OCS looks relatively inert, except for the last part of the simulation ($t > 10^5$ years). Indeed, in the 0DS100EDC case, SO and SO$_2$ (cf figure \ref{fig_3} (d) and (e)) are at first slowly formed through reactions (\ref{eq_9}) and (\ref{eq_18}), respectively. As the abundance of HCNS grows from the evaporated NS (accumulated on the grain during the dense parent cloud phase, see table \ref{tab_2}) through:

 \begin{eqnarray}
 	\text{NS} + \text{CH$_2$} &\to &\text{HCNS} + \text{H} 
	\label{eq_22}
\end{eqnarray}

the abundance of SO grows steeper from 10$^4$ years thanks to reaction (\ref{eq_4a}). Finally, after 10$^6$ years, SO is consumed by CH through reactions (\ref{eq_17a}) and (\ref{eq_17b}). The strong link of SO$_2$ with SO via reaction (\ref{eq_18}), coupled with reaction (\ref{eq_19a}), makes its abundance drops at the same time. Regarding OCS (cf figure \ref{fig_3} (f)), its abundance grows at first in the 0DS100EDC case from reaction (\ref{eq_15a}), then from reaction (\ref{eq_20b}) after 10$^4$ years. Towards the end of the simulation, as for SO and SO$_2$, OCS is destroyed by CH via:

 \begin{eqnarray}
 	\text{OCS} + \text{CH} &\to &\text{H} + \text{CO} + \text{CS}
	\label{eq_25}
\end{eqnarray}\\

In the 0DS300EDC case, the chemistry of SO, SO$_2$ and OCS differs from the 0DS100EDC case from two main points: first, the abundance of OH decreases much faster in the 0DS300EDC case (cf figure \ref{fig_2} (f)), diminishing even more the quantity of reactive oxygen in the gas phase, causing the abundances of SO, SO$_2$ and OCS to stay relatively constant during at least the first 10$^5$ years of the simulation. Secondly, CH is much less abundant at 300 K (by two to three orders of magnitude) because it is effectively destroyed by evaporated H$_2$O and C$_2$H$_2$:

 \begin{eqnarray}
 	\text{CH} + \text{H$_2$O} &\to &\text{H} + \text{H$_2$CO}\\
	\text{CH} + \text{C$_2$H$_2$} &\to &\text{H} + \text{c-C$_3$H$_2$}
	\label{eq_26}
\end{eqnarray}

Hence, SO and SO$_2$ abundances keep on increasing at the end of the simulation (cf figure \ref{fig_3} (d) and (e)) and do not decrease as in the 0DS100EDC case. However, the lack of CH do not prevent the late-time decrease of OCS (cf figure \ref{fig_3} (f)) which is ruled by the same reactions as in the 0DS300LEDC case, namely its photodissociation by secondary UV photon and reaction (\ref{eq_21}).\\~\\

\begin{figure*}
        \begin{center}
                \includegraphics[scale=0.13]{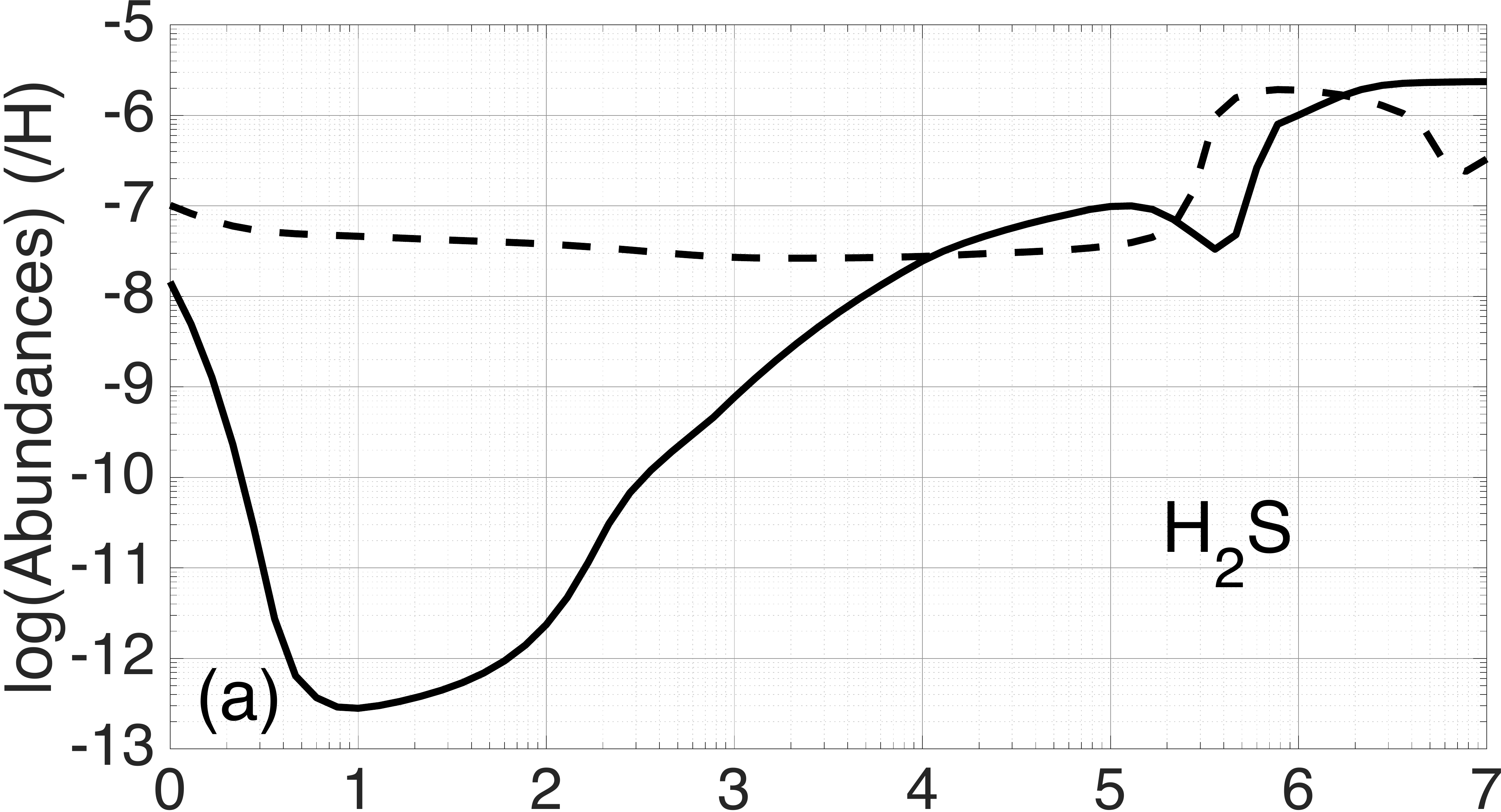}
                \includegraphics[scale=0.13]{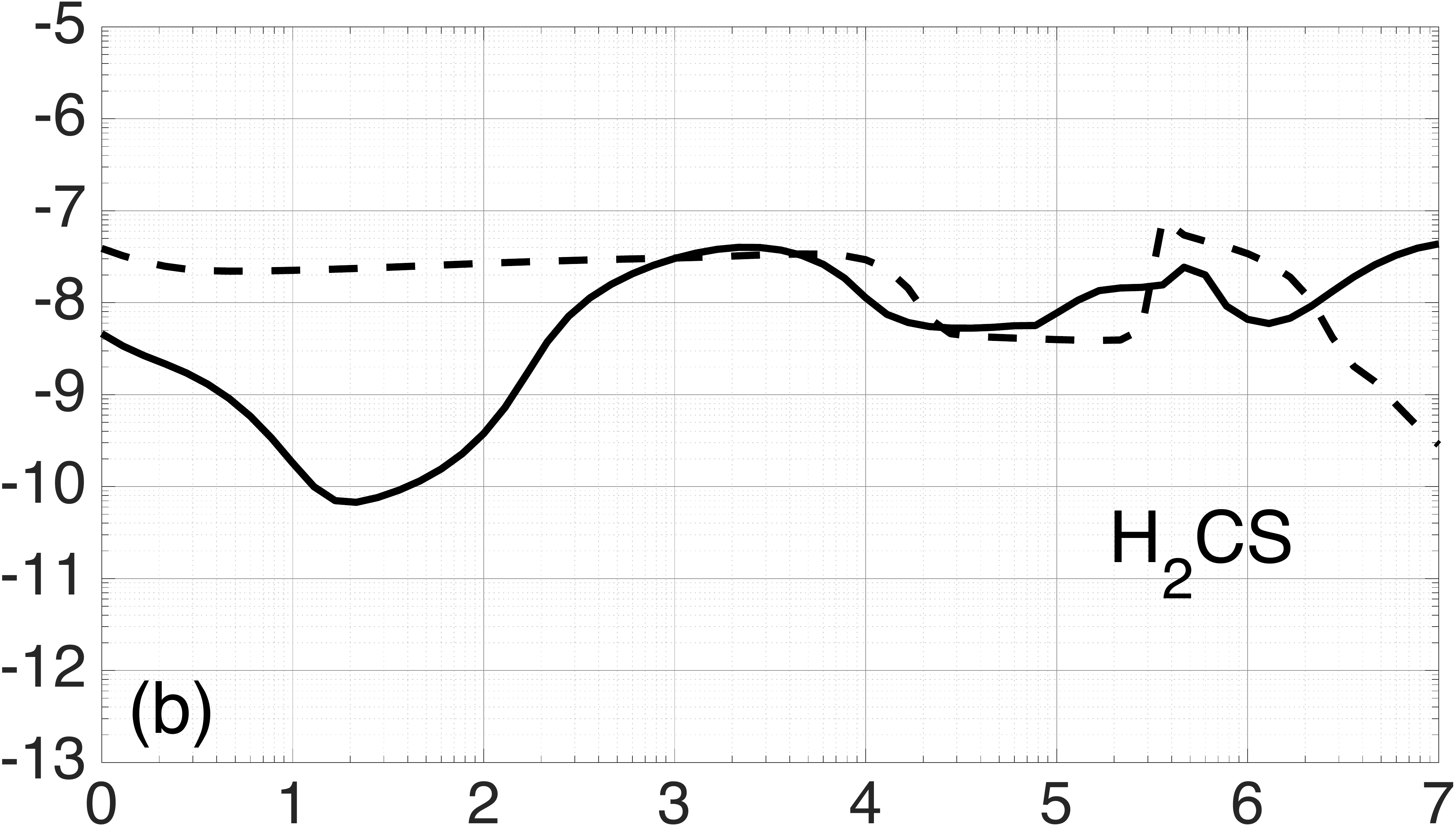}
                \includegraphics[scale=0.13]{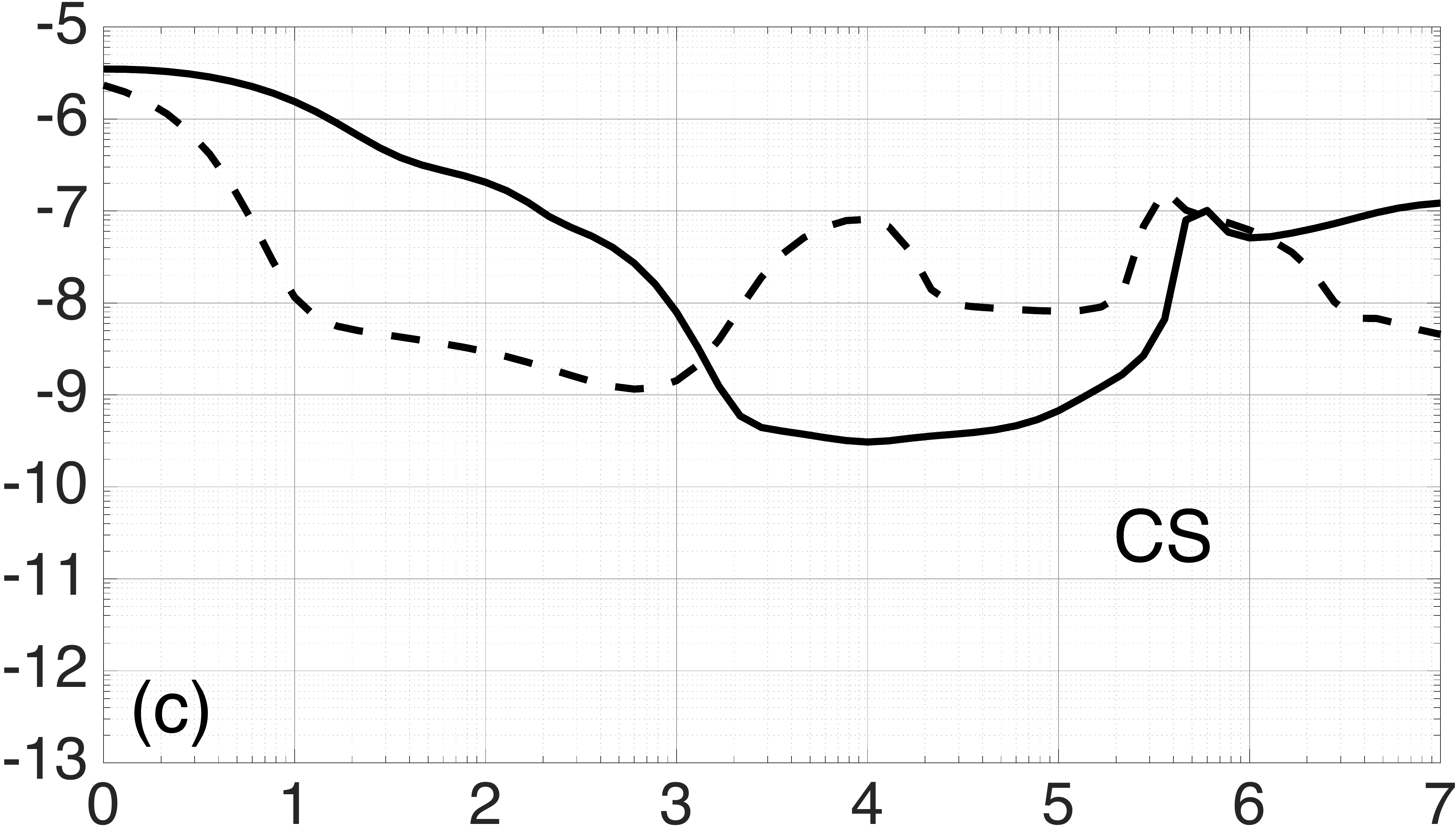}\\
                \includegraphics[scale=0.13]{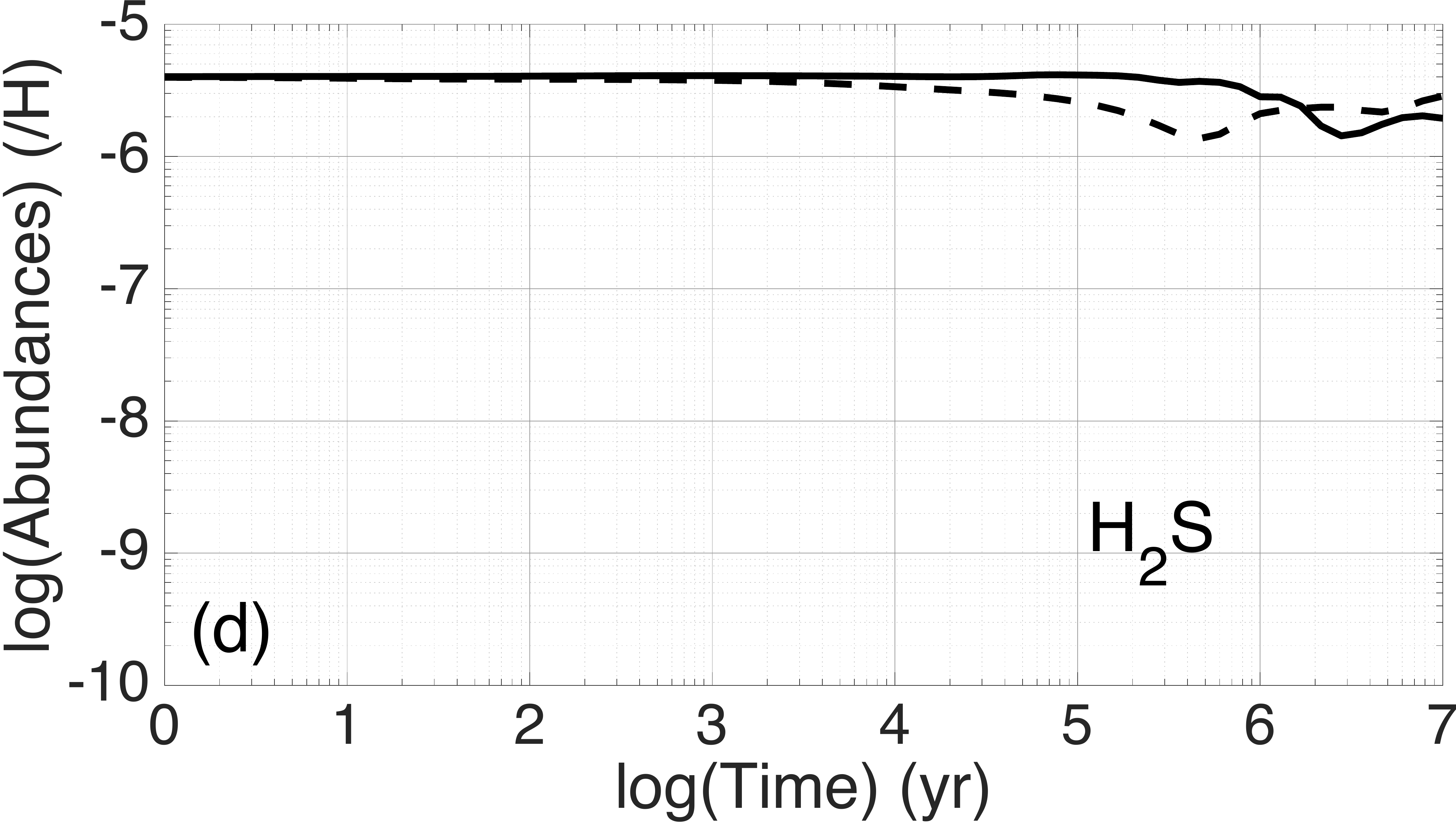}
                \includegraphics[scale=0.13]{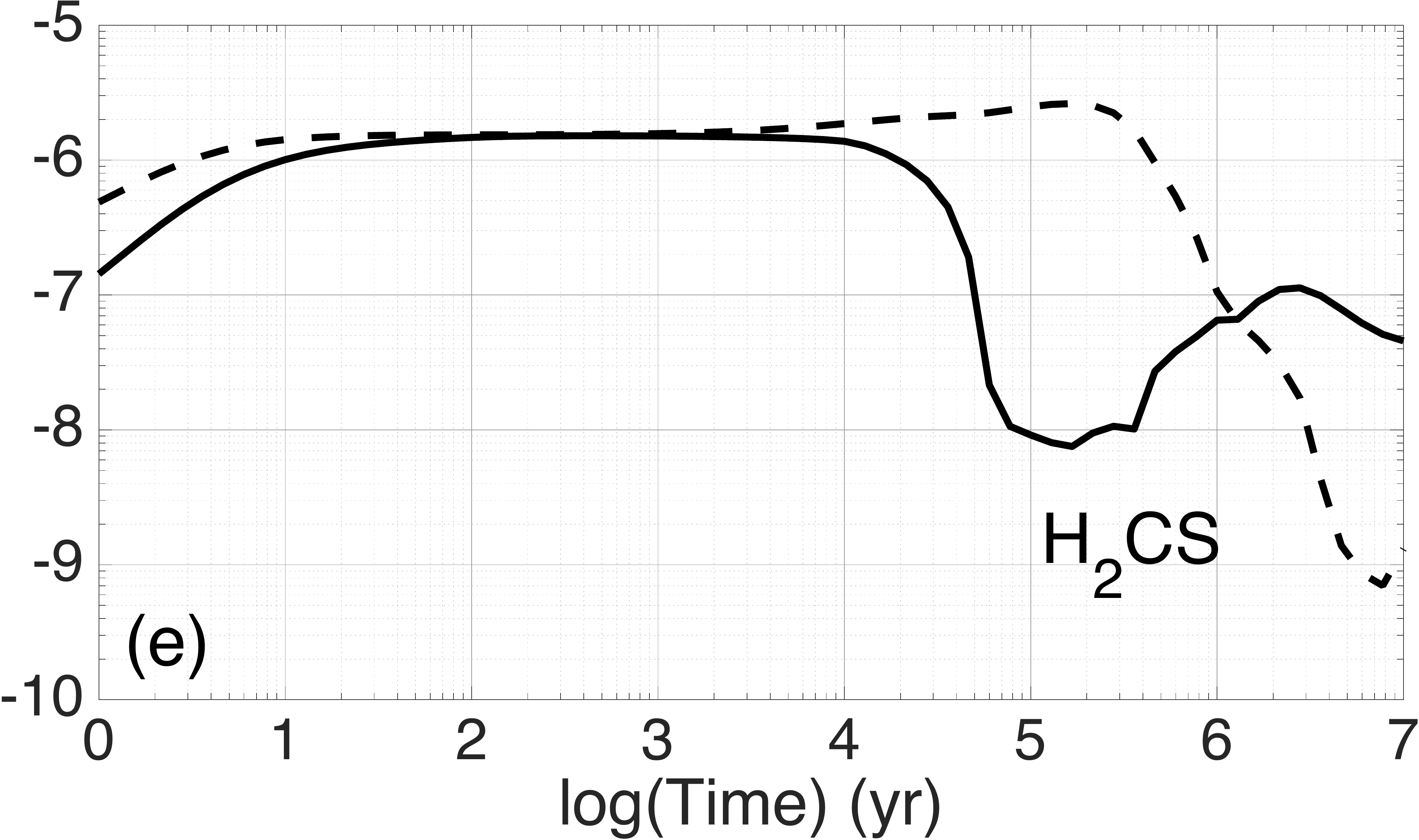}
                \includegraphics[scale=0.13]{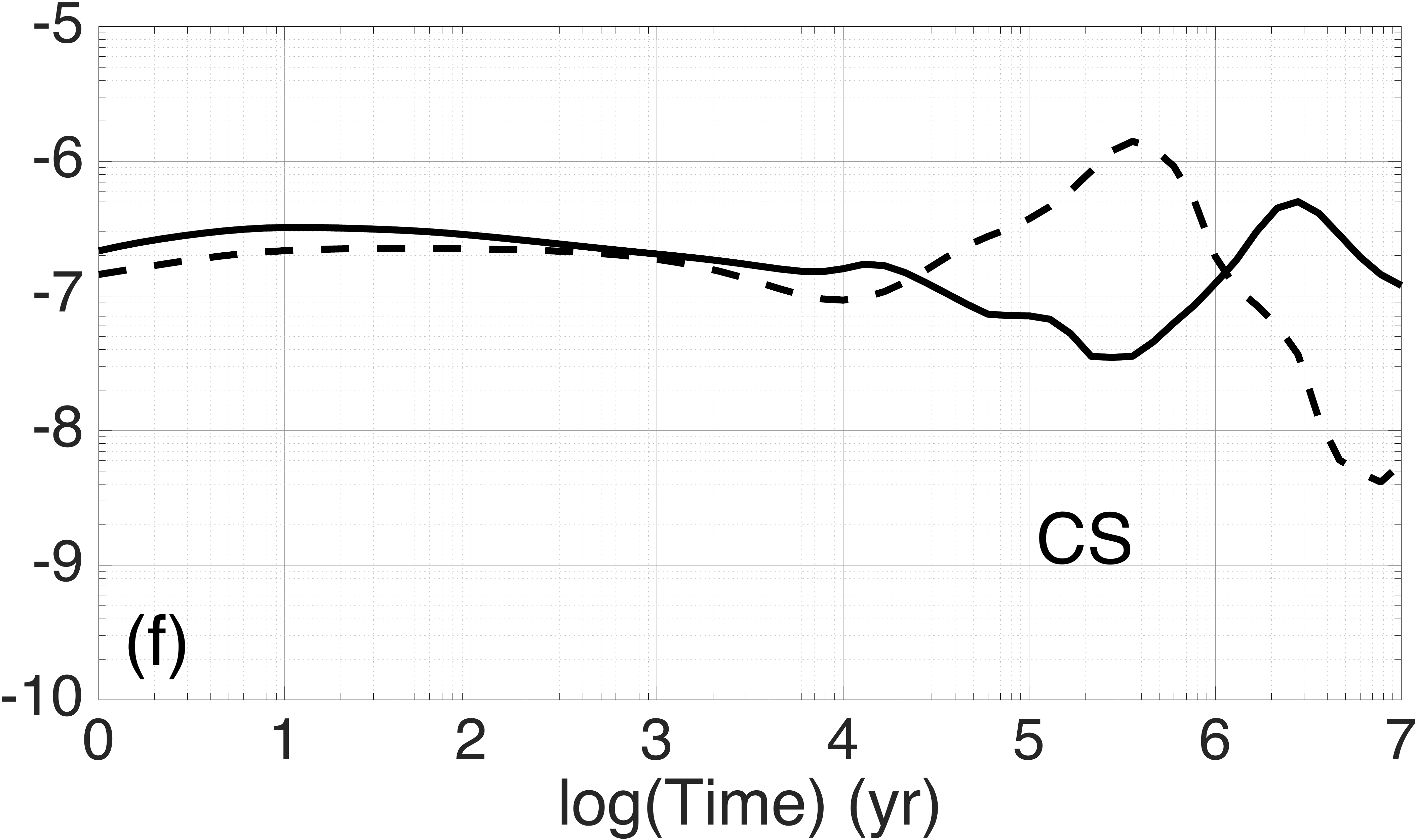}
                \caption{Abundances of H$_2$S, H$_2$CS and CS relative to H as a function of time for hot cores conditions: $n_H = 2\times10^7$ cm$^{-3}$ and T = 100 K (solid line) or 300 K (dashed line), and for LEDC (top panel) and EDC (bottom panel) pre-collapse compositions.}
                \label{fig_4}
        \end{center}
\end{figure*}

3.1.2.3 \quad H$_2$S, H$_2$CS and CS in the LEDC cases\\

Figure \ref{fig_4} is the same as figure \ref{fig_3} but for H$_2$S, H$_2$CS and CS. In the 0DS100LEDC case, as for SO, SO$_2$ and OCS, H$_2$S and H$_2$CS (cf figure \ref{fig_4} (a) and (b)) are both initially destroyed by atomic carbon via the following reactions:

 \begin{eqnarray}
 	\text{H$_2$S} + \text{C} &\to &\text{H} + \text{HCS} \label{eq_27a}\\
	\text{H$_2$CS} + \text{C} &\to &\text{H} + \text{HC$_2$S}\\
	&\to &\text{CS} + \text{CH$_2$}
	\label{eq_27c}
\end{eqnarray}

The following increase in H$_2$S abundance is mainly due to two coupled reactions:

 \begin{eqnarray}
 	\text{CH$_2$SH} + \text{N} &\to &\text{H$_2$S} + \text{HCN} 	\label{eq_28a}\\
	\text{HCN} + \text{H$_3$S$^+$} &\to &\text{H$_2$S} + \text{HCNH$^+$}
	\label{eq_28b}
\end{eqnarray}

Indeed, CH$_2$SH evaporates slowly from grain surface in the physical conditions of the simulation. Reacting with atomic nitrogen via reaction (\ref{eq_28a}), it forms both H$_2$S and HCN. The latter then also forms H$_2$S. The combination of these two reactions explains the steep increase in H$_2$S abundance between 10 and 10$^5$ years. Afterwards, it decreases from the following ion-neutral reactions:

 \begin{eqnarray}
 	\text{H$_2$S} + \text{HCO$^+$} &\to &\text{CO} + \text{H$_3$S$^+$}\label{eq_29a}\\
	\text{H$_2$S} + \text{SO$^+$} &\to &\text{H$_2$O} + \text{S$_2^+$}
	\label{eq_29b}
\end{eqnarray}

Towards the end of the simulation, H$_2$S is efficiently produced by the electronic recombination of H$_3$S$_2^+$, provided by the following reaction mechanism:

\begin{eqnarray}
	\text{S} + \text{HS} \to \text{S$_2$} \xrightarrow{\text{HCO$^+$}} \text{HS$_2^+$} \xrightarrow{\text{H$_2$}} \text{H$_3$S$_2^+$}
	\label{rm_1}
\end{eqnarray}

This mechanism is made efficient by the increase of the abundance of HS during this period of time.\\
H$_2$CS gas-phase chemistry at 100 K (in both LEDC and EDC cases, cf figure \ref{fig_4} (b) and (e), respectively) is intertwined with its grain chemistry because the temperature is not high enough for its complete thermal desorption from grain ices. Hence, after its destruction by atomic carbon, H$_2$CS grows from:

 \begin{eqnarray}
 	\text{S} + \text{CH$_3$} &\to &\text{H$_2$CS} + \text{H} 
	\label{eq_30}
\end{eqnarray}

Then its abundance decreases from destruction by HCNH$^+$ in the gas phase, as well as by hydrogenation in the grain bulk. Indeed the latter contributes to the depletion of H$_2$CS from the gas phase at this time because the chemistry has reached adsorption/desorption equilibrium. The reactions in question are respectively:

 \begin{eqnarray}
 	\text{H$_2$CS} + \text{HCNH$^+$} &\to & \text{H$_3$CS$^+$} + \text{HCN} \label{eq_31a}\\
	\text{b-H$_2$CS} + \text{b-H} &\to & \text{b-CH$_2$SH}
	\label{eq_31b}
\end{eqnarray}

where the prefix 'b-' is for the bulk species. Afterwards, H$_2$CS abundance increases again, mainly from the electronic recombination of H$_3$CS$^+$ and reaction (\ref{eq_30}).\\
Contrary to the other S-bearing species studied in this paper, CS is not destroyed by atomic carbon during the early phase of the simulation (cf figure \ref{fig_4} (c)). It is instead destroyed during a longer period of time (approximately 10$^4$ years) by atomic oxygen, contributing efficiently to lock gas phase oxygen into CO via:

 \begin{eqnarray}
 	\text{CS} + \text{O} &\to &\text{S} + \text{CO} 
	\label{eq_32}
\end{eqnarray}

CS abundance then grows mainly from reaction (\ref{eq_16b}) as well as from the following reaction mechanism:

\begin{eqnarray}
	\text{S} + \text{CH$_2$} \to \text{HCS} \xrightarrow{\text{S}} \text{CS$_2$} \xrightarrow{\text{H}} \text{CS}
	\label{rm_2}
\end{eqnarray}\\

As previously observed, atomic carbon is preferentially consumed by evaporated hydrocarbon in the 0DS300LEDC case and therefore has a limited impact on sulphur chemistry compared to the 0DS100LEDC case. Hence, H$_2$S (cf figure \ref{fig_4} (a)) is first destroyed by atomic hydrogen and oxygen for 10$^3$ years via:

 \begin{eqnarray}
 	\text{H$_2$S} + \text{H} &\to &\text{H$_2$} + \text{HS}\label{eq_33a}\\
	\text{H$_2$S} + \text{O} &\to &\text{H$_2$} + \text{OH}
	\label{eq_33b}
\end{eqnarray}

H$_2$S is after formed by the electronic recombination of H$_3$S$_2^+$ provided by a similar reaction mechanism than (\ref{rm_1}), except that instead of HCO$^+$, it is the H$_3$O$^+$ ion which mainly reacts with S$_2$ to form HS$_2^+$. Indeed H$_3$O$^+$ is much more abundant in the gas phase at 300 K than at 100 K because it comes mainly from H$_2$O which is totally evaporated from grain at this temperature. Furthermore, it is H$_3$O$^+$ which destroys H$_2$S at the end of the simulation:

 \begin{eqnarray}
 	\text{H$_2$S} + \text{H$_3$O$^+$} &\to &\text{H$_2$O} + \text{H$_3$S$^+$} 
	\label{eq_34}
\end{eqnarray}

At 300 K, H$_2$CS is fully depleted from grain surface and bulk and is affected only by gas phase chemistry (cf figure \ref{fig_4} (b)). Partially consumed by atomic carbon during the first 10 years, it is then formed mainly through reaction (\ref{eq_30}) and destroyed by reaction (\ref{eq_31a}).\\
In the 0DS300LEDC case, CS is also initially destroyed by atomic oxygen via reaction (\ref{eq_32}) (cf figure \ref{fig_4} (c)), but for a shorter time than in the 0DS100LEDC case. The increase in its abundance at 10$^3$ years is mostly due to a reaction mechanism starting from the evaporated C$_4$H$_2$:

\begin{eqnarray}
	\text{C$_4$H$_2$} + \text{S$^+$} \to \text{HC$_4$S$^+$} \xrightarrow{\text{e$^-$}} \text{C$_3$S} \xrightarrow{\text{O}} \text{CCS} \xrightarrow{\text{S,N}} \text{CS}
\end{eqnarray}

Then CS is mainly destroyed by HCNH$^+$ via: 

 \begin{eqnarray}
 	\text{CS} + \text{HCNH$^+$} &\to & \text{HCN} + \text{HCS$^+$} 
	\label{eq_35}
\end{eqnarray}

Finally, the last increase in its abundance is due to the reaction mechanism (\ref{rm_2}).\\~\\

3.1.2.4 \quad H$_2$S, H$_2$CS and CS in the EDC cases\\

In the EDC cases and prior to collapse, H$_2$S in icy grain bulk is the second reservoir of sulphur, containing 26\% of the total amount of sulphur (see table \ref{tab_2}). Hence its abundance in both EDC cases is generally higher than in the LEDC ones. Moreover, in the 0DS100EDC case (cf figure \ref{fig_4} (d)), H$_2$S initially forms efficiently from its abundant reservoirs counterpart HS via:

 \begin{eqnarray}
 	\text{HS} + \text{HS} &\to & \text{S} + \text{H$_2$S} 
	\label{eq_36}
\end{eqnarray}

It is also produced during most of the simulation from the slow evaporation of methanol, which is the fourth reservoir of oxygen in the EDC case (see table \ref{tab_2}), and through the following reaction mechanism:

\begin{eqnarray}
	\text{CH$_3$OH} + \text{CN} \to \text{CH$_2$OH} \xrightarrow{\text{S}} \text{H$_2$CO} \xrightarrow{\text{H$_3$S$^+$}} \text{H$_2$S} 
	\label{rm_3}
\end{eqnarray}

Towards the end, H$_2$S is destroyed via reaction (\ref{eq_29a}).\\
In the 0DS100EDC case (cf figure \ref{fig_4} (e)), H$_2$CS is first formed via reaction (\ref{eq_30}) as well as from methanol slow depletion with the following reaction mechanism, similar to (\ref{rm_3}):

\begin{eqnarray}
	\text{CH$_3$OH} + \text{CN} \to \text{CH$_2$OH} \xrightarrow{\text{S}} \text{H$_2$CS}	
	\label{rm_4}
\end{eqnarray}

As in the 0DS100LEDC case, H$_2$CS chemistry is afterwards linked with its grain chemistry via reactions (\ref{eq_31a}) and (\ref{eq_31b}), making its abundance drop at 10$^5$ years. At the end of the simulation H$_2$CS abundance increases again from reaction (\ref{eq_30}) and electronic recombination of H$_3$CS$^+$.\\
Due to the small amount of reactive oxygen in the EDC cases compared to the LEDC cases (see table \ref{tab_4}), CS is not destroyed by atomic oxygen at the beginning of the simulation (cf figure \ref{fig_4} (e)). At 100K, it is instead formed rapidly via:

 \begin{eqnarray}
 	\text{HCS} + \text{H} &\to & \text{CS} + \text{H$_2$} 
	\label{eq_37}
\end{eqnarray}
	
with HCS coming from the small fraction of H$_2$S destroyed by reaction (\ref{eq_27a}). Afterward, CS is destroyed for a few 10$^5$ years by OH and HS via reaction \ref{eq_15a} and:

 \begin{eqnarray}
 	\text{CS} + \text{HS} &\to & \text{CS$_2$} + \text{H} \label{eq_38a}
\end{eqnarray}

When CS$_2$ abundance is high enough, it is hydrogenated back into CS, causing the abundance of the latter to grow again at the end of the simulation.\\

In the 0DS300EDC case, all the methanol is directly depleted in the gas phase and rapidly forms CH$_3$ through many different reactions. Hence, as CH$_3$ abundance is rapidly much higher than in the 0DS100EDC case, H$_2$S is mainly destroyed at 300K via (cf figure \ref{fig_4} (d)):

 \begin{eqnarray}
 	\text{H$_2$S} + \text{CH$_3$} &\to & \text{HS} + \text{CH$_4$} 
	\label{eq_39}
\end{eqnarray}
	
This high abundance of CH$_3$ in the gas phase also causes H$_2$CS to be efficiently formed for a few 10$^5$ years by reaction (\ref{eq_30}) (cf figure \ref{fig_4} (e)). Afterwards, as in the 0DS300LEDC case, the steep decrease of its abundance is mainly due to reaction (\ref{eq_31a}).\\
Finally, CS chemistry in the 0DS300EDC case is similar to the 0DS100EDC case, except for the fact that OH abundance at 300K is much lower than at 100K (cf figure \ref{fig_4} (f)). Hence, CS is mainly destroyed by reaction (\ref{eq_38a}). Moreover, as CS$_2$ abundance grows faster at 300K than at 100K, the peak in CS abundance happens sooner in the 0DS300LEDC case. At the end, CS is destroyed like H$_2$CS, by HCNH$^+$ via reaction (\ref{eq_35}).

\subsubsection{Comparisons to observations}

The species studied in this section have been detected in many hot cores and corinos, and their respective observed abundances present differences among sources that can go as high as three orders of magnitude \citep[see for example table 5 of][and reference therein]{Wakelam04b}. These variations are often explained by differences among the ages of the sources, or among the temperatures of their respectives hot cores or corinos \citep[see discussion in][]{Herpin09}. Therefore, it would be complex, as well as out of the scope of the present study, to quantitatively compare our results to observations. Qualitatively however, we can raise the two following points:\\
\begin{enumerate}
\item The total amount of sulfur observed in massive hot cores generally account only for a small part of its cosmic abundance \citep[around 0.1\%, see][]{Hatchell98,VDTak04,Wakelam04b,Herpin09}, which contrasts with our modeling results where most of the sulphur appears to be under the form of SO, SO$_2$, H$_2$S and OCS at the ages that are expected for such objects. Uncertainties on massive hot cores observations due to the fact they are mostly very distant sources, therefore not spatially resolved, as well as uncertainties on our high temperature network could explain these discrepancies. However, among this type of sources, the hot core of Orion KL presents a high abundance of H$_2$S of $2.5\times10^{-6}$, accounting for more than 15\% of the total mount of sulphur \citep{Minh90}, as well as higher abundances of SO and SO$_2$ than in other massive hot core \citep{Sutton95}. It appears that, even if our models fail to reproduce the observed abundances of these molecules for most massive hot cores observations, both our EDC case models can reproduce the Orion KL abundances of H$_2$S, SO and SO$_2$ within one order of magnitude in a range of time acceptable for this structure, between 10$^4$ and 10$^6$ years. However in this range of time, the models tend to overestimate the abundances of OCS, CS and H$_2$CS, which suggests that work still has to be done regarding the modelling of the chemistry of these species.\\
\item The only hot corinos towards which all the S-bearing species studied in this paper have been observed is IRAS 16293-2422. The observed abundances of SO, SO$_2$, OCS, H$_2$CS and H$_2$S in the dense inner part of its envelope \citep[$\le150$ AU, see table 7 of][]{Schoier02} can be reproduced within one order of magnitude by both our LEDC case models in a range of time compatible with one derive by \citet{Schoier02}, between a few 10$^3$ and a few 10$^4$ years. This result would suggest that IRAS 16293-2422 has formed in a parent cloud that would have collapsed at an age of approximately 10$^5$ years.
\end{enumerate} 

These results suggest that, following our previous paper \citep{Vidal17}, our model can reproduce observations of S-bearing molecules in H$_2$S-rich hot cores and in hot corinos using as initial abundance of sulphur its cosmic one. However work still has to be conducted regarding our high temperature network, especially for OCS, H$_2$CS and CS.

\subsection{1D static models} \label{sec_1D}

In this section we study the results of the two 1D static simulations whose parameters are described in section \ref{sec_1Dp} and table \ref{tab_3}. The goal is to evaluate the impact of the pre-collapse chemical composition of the parent cloud on the computed abundances of the main S-bearing species SO, SO$_2$, OCS, H$_2$S, H$_2$CS and CS. Both LEDC and EDC simulations were run for a period of $3.5\times10^5$ years so as to make the results comparable to those of the 0D dynamic simulations presented in section \ref{sec_dyn}.\\

\begin{figure*}
        \begin{center}
                \includegraphics[scale=0.13]{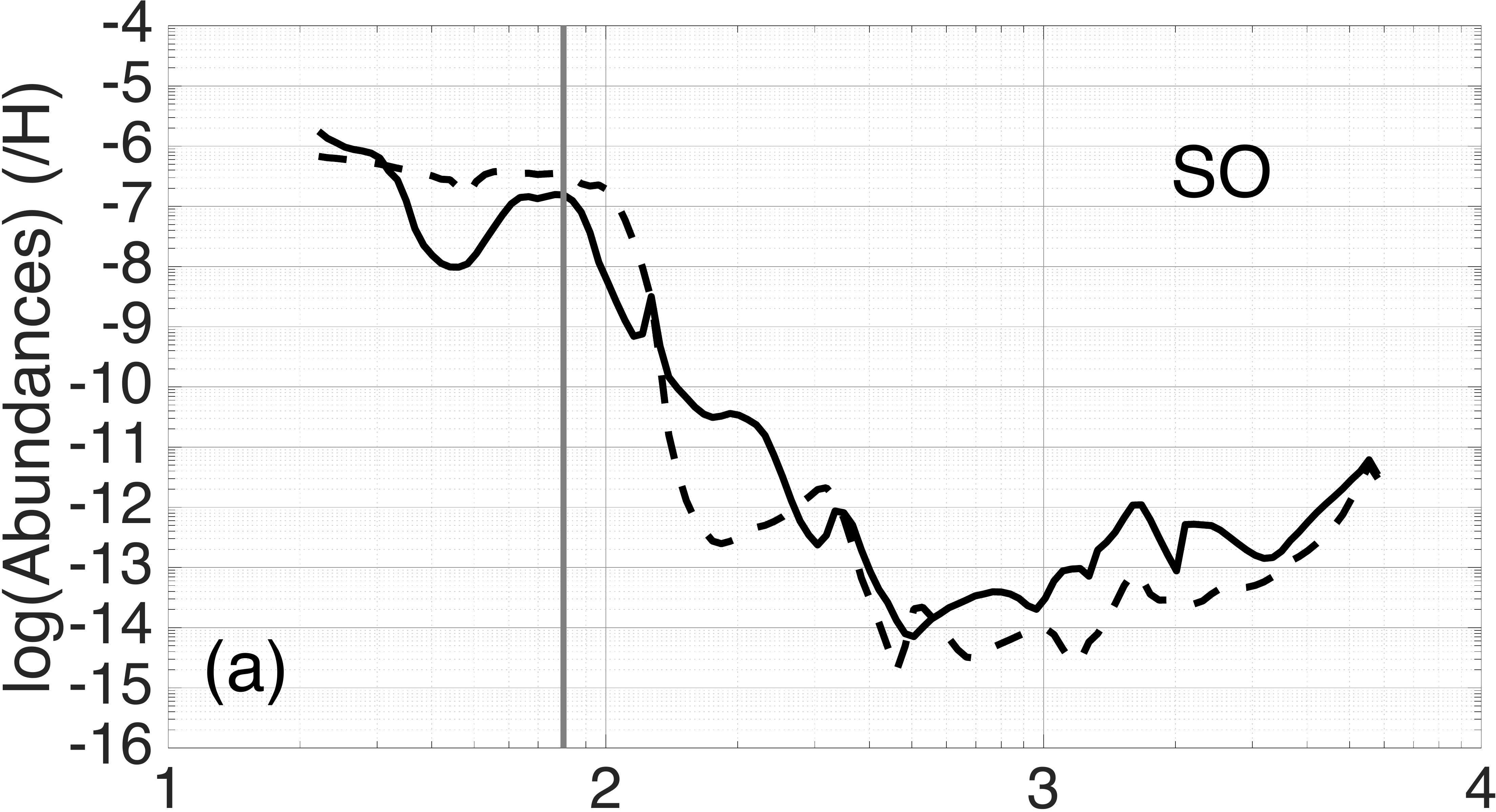}
                \includegraphics[scale=0.13]{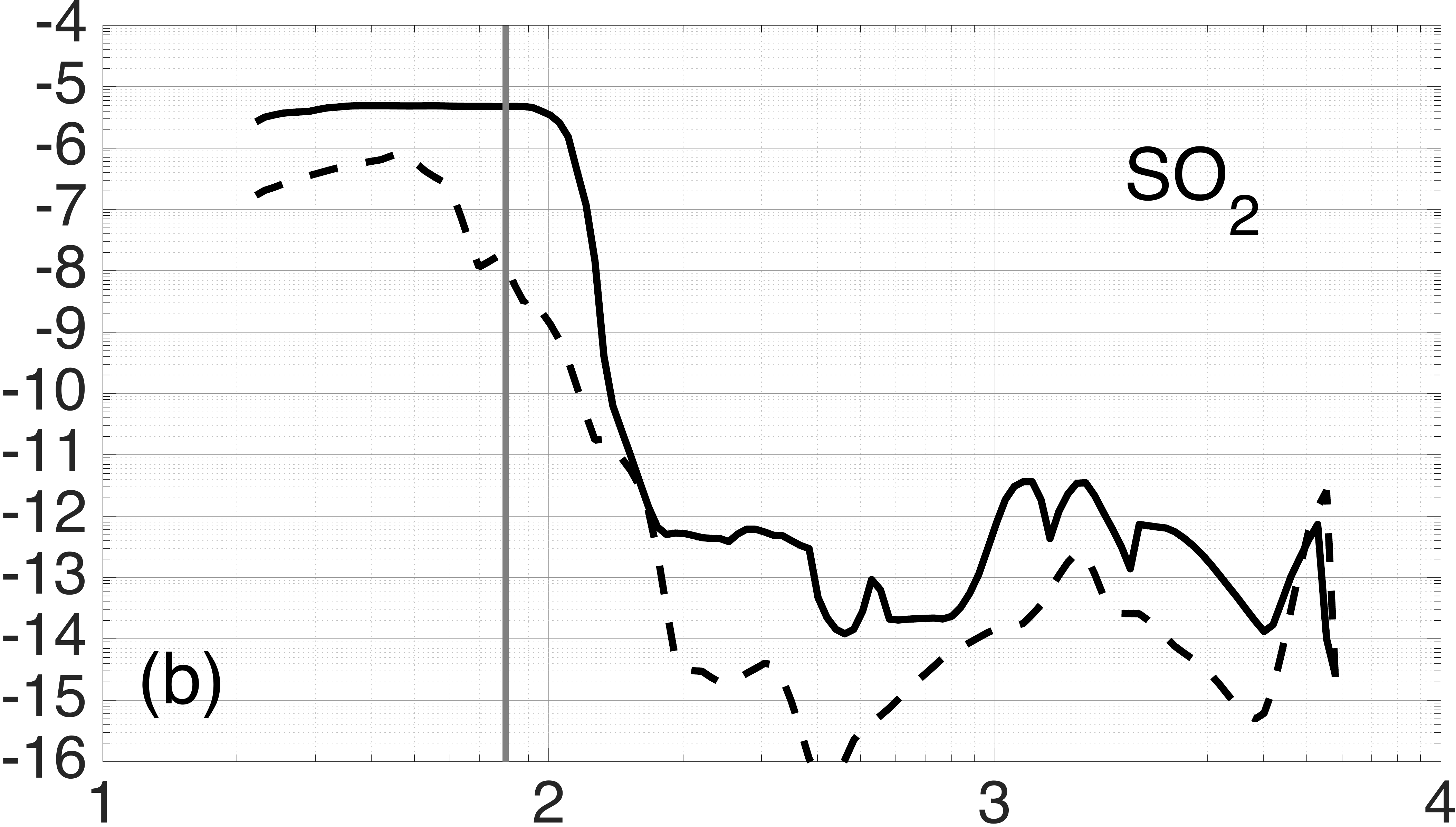}
                \includegraphics[scale=0.13]{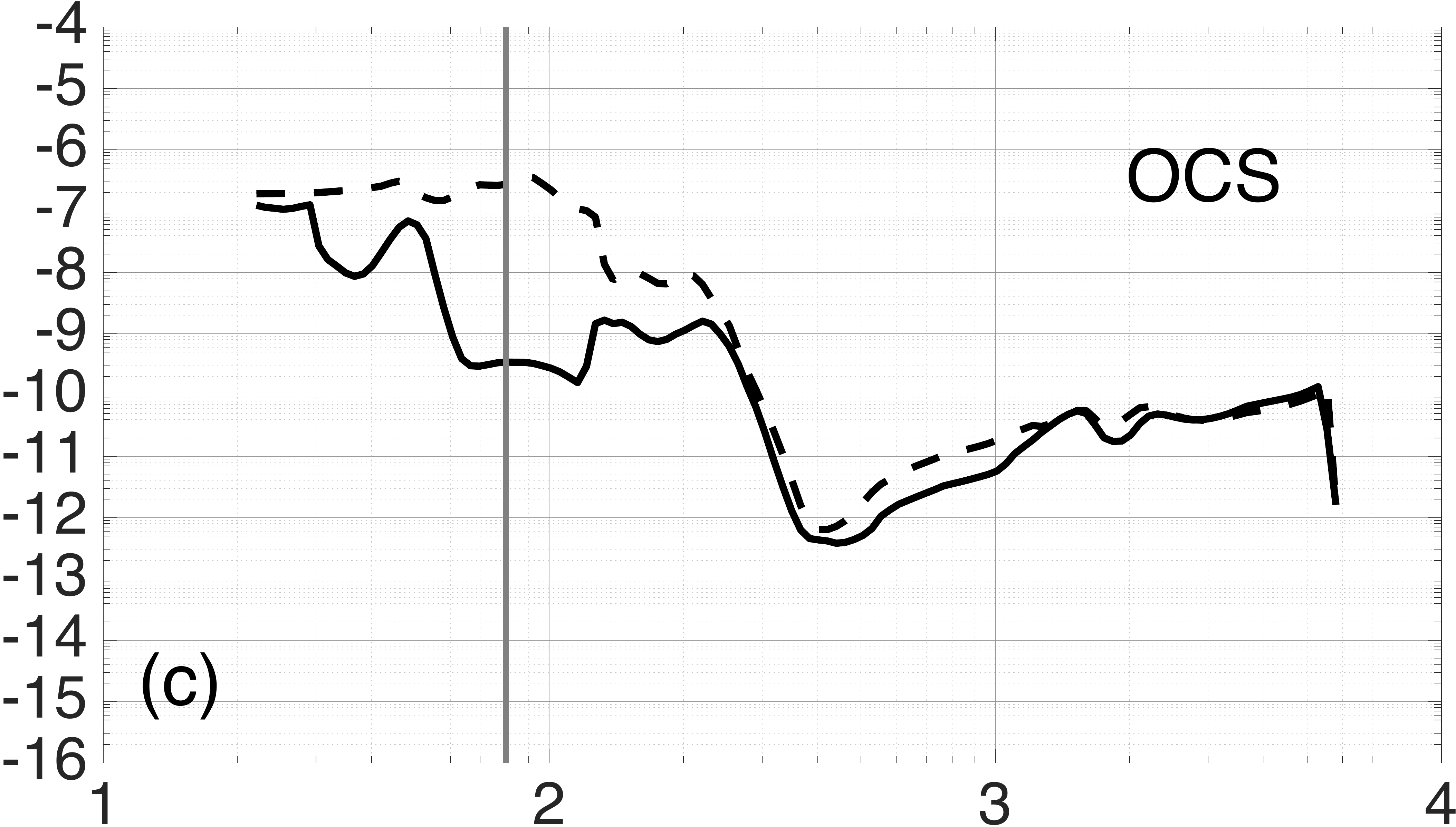}\\
                \includegraphics[scale=0.13]{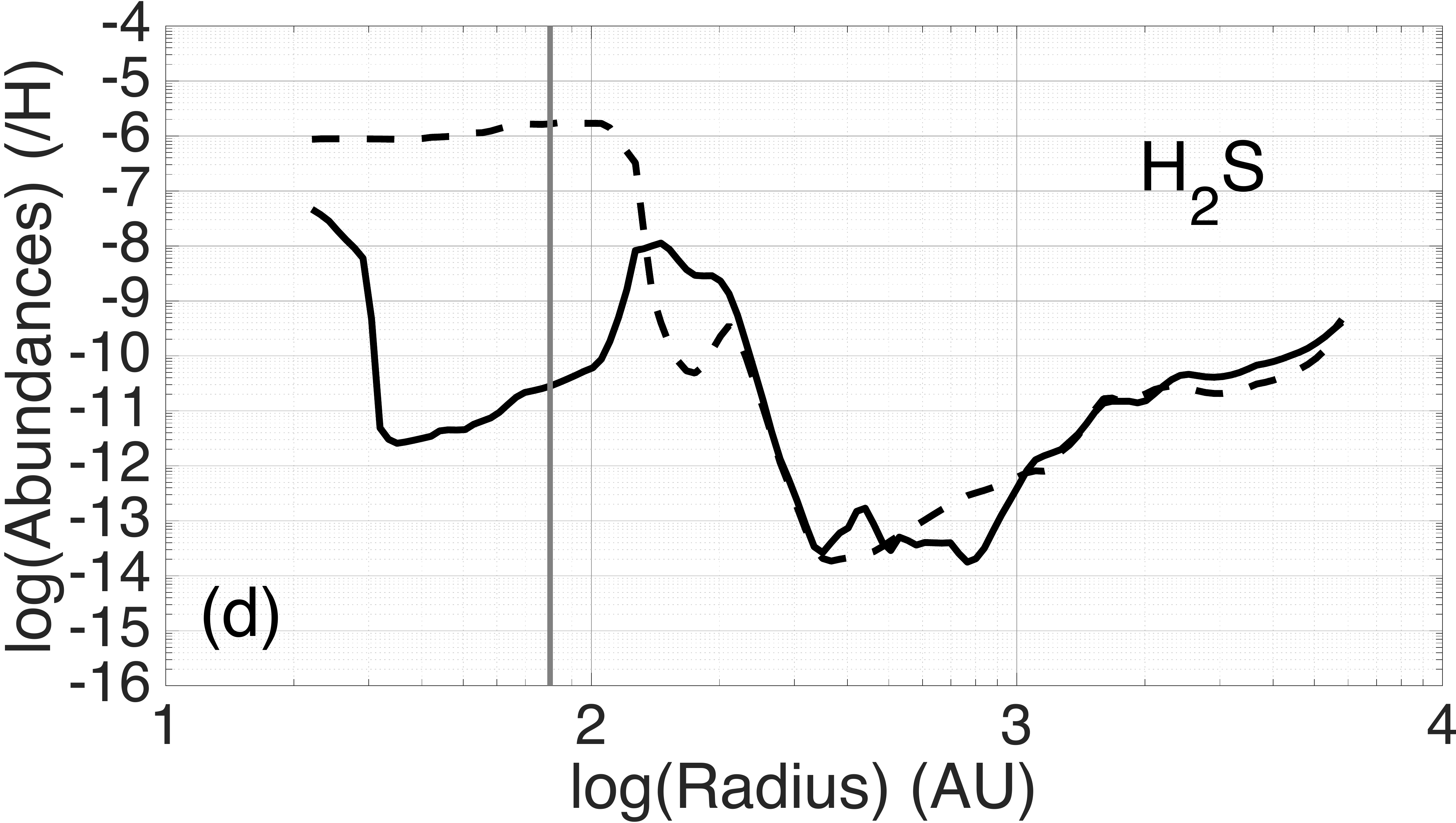}
                \includegraphics[scale=0.13]{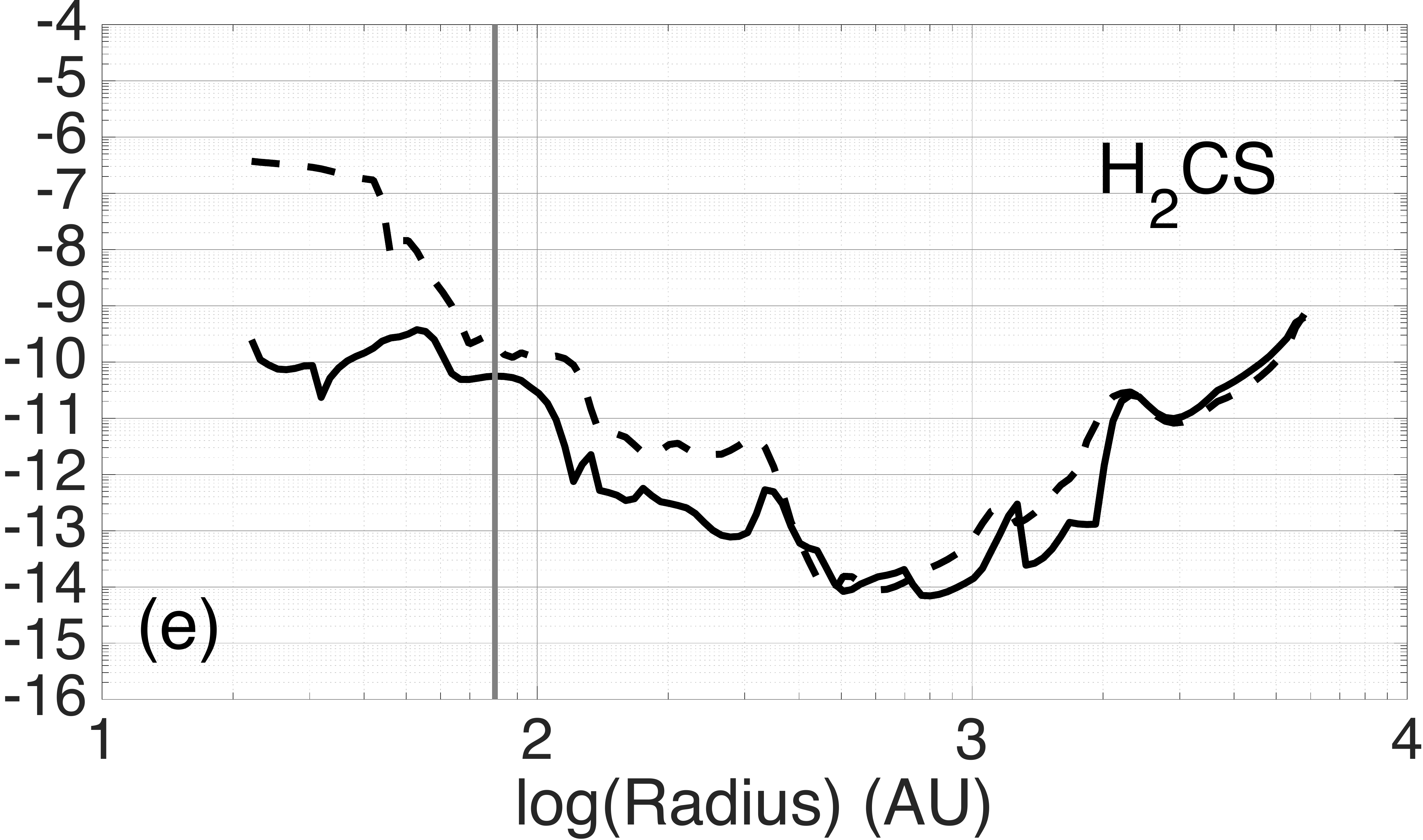}
                \includegraphics[scale=0.13]{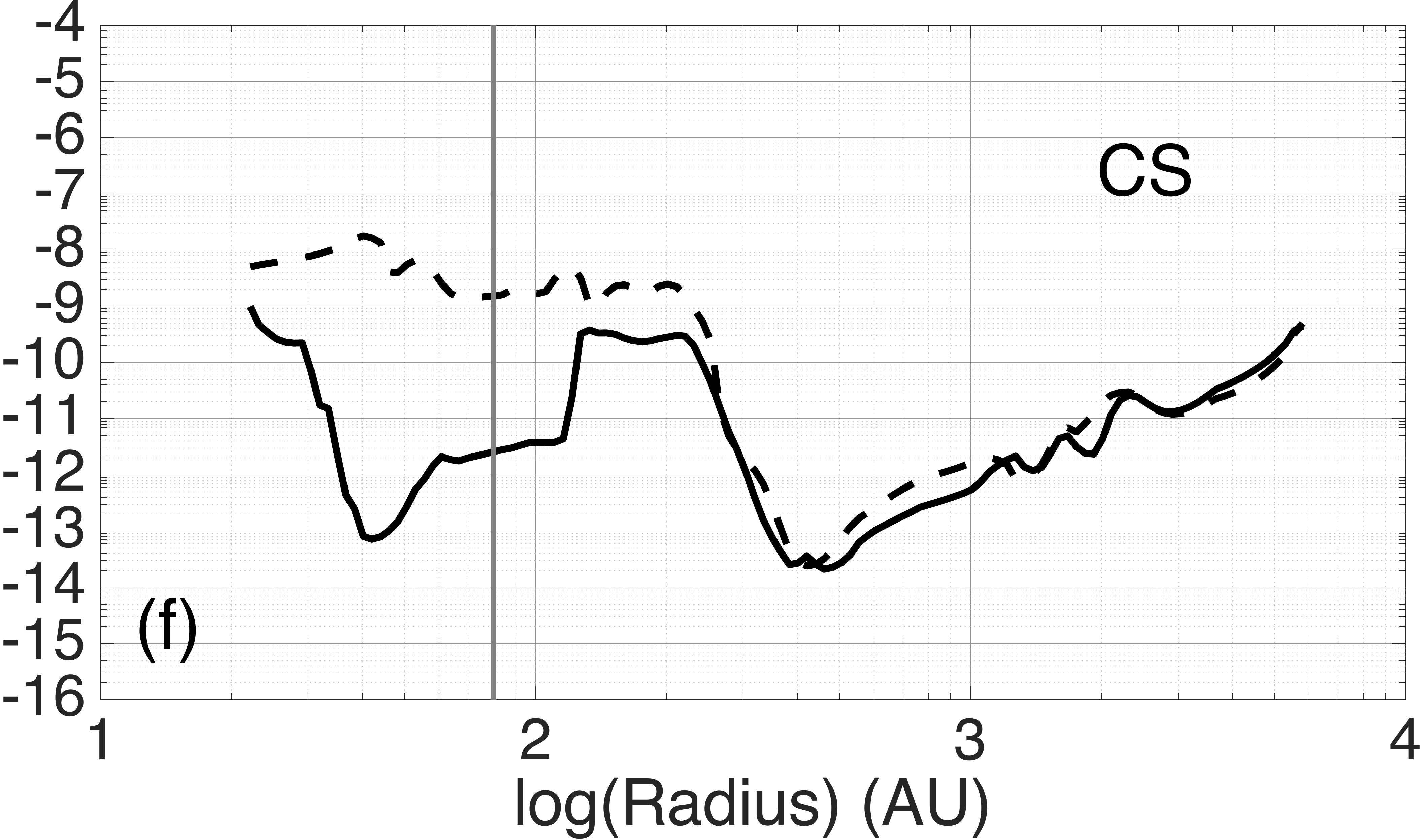}
                \caption{Abundances of SO, SO$_2$ and OCS (top panel) and H$_2$S, H$_2$CS and CS (bottom panel), relative to H as a function of the radius to the star IRAS 16293-2422 according to the 1D structure of \citet{Crimier10}, for the LEDC (solid line) and EDC (dashed line) pre-collapse compositions. Both models were run for a period of $3.5\times10^5$ years so the results would be comparable to those of the 0D dynamic simulations. The vertical grey line represent the hot core spatial limit $R_{HC}$ = 80 AU, T > 100 K.}
                \label{fig_5}
        \end{center}
\end{figure*}

Figure \ref{fig_5} displays the abundances of each of these species for both 1DSLEDC (solid lines) and 1DSEDC (dashed lines) pre-collapse compositions. For a radius greater than 300 AU, which corresponds to the outermost and coldest (T < 50 K) part of the envelope of the protostar, most considered species present only small local differences in their abundances between the LEDC and EDC cases. Only SO$_2$ presents significant differences that can go to more than three orders of magnitude while other species display differences smaller than one order of magnitude (cf figure \ref{fig_5} (b)).\\

In the inner part of the envelope (R < 300 AU) where the temperature goes from 50 to 200 K (see right panel of figure \ref{fig_1}), all species abundances show significative differences between the LEDC and EDC pre-collapse compositions. These differences can go from two to more than six orders of magnitude. Within the hot core limits (delimited by a vertical grey line in figure \ref{fig_5} at $R_{HC}$ < 80 AU, T > 100 K), the species which are the most sensitive to the pre-collapse composition appear to be H$_2$S, H$_2$CS, and CS with differences of more than three orders of magnitude (cf figure \ref{fig_5} (d), (e) and (f), respectively). For H$_2$S, these differences are explained by its dependance on the parent cloud evolution time. Indeed, in the EDC case, H$_2$S in icy grain bulk is the second reservoirs of sulphur in the pre-collapse composition, containing 26\% of the total amount of sulphur, whereas in the LEDC case, it only contains 5\% (see table \ref{tab_2}). Hence, in the inner part of the envelope and the hot core, where the temperature is high enough for H$_2$S thermal desorption, its abundance is much higher than in the LEDC case. Moreover, figure \ref{fig_4} shows that in that case, H$_2$S is not efficiently destroyed in the gas phase. Regarding H$_2$CS, this species efficiently forms at high temperature in the gas phase from CH$_3$ (via reaction (\ref{eq_30})) which is much more abundant in the EDC case because of evaporated methanol and hydrocarbons accumulated on grain during the parent cloud evolution. Finally, in the EDC case CS is not as efficiently destroyed in the gas phase as in the LEDC case (see figure \ref{fig_4}) because of the low abundance of reactive oxygen (see table \ref{tab_4}).\\

We can finally highlight that our 1D static models show that the pre-collapse composition of the parent cloud appears to be critical for the sulphur bearing species in hot core physical conditions. Indeed, we can see from figure \ref{fig_5} that a hot core that formed from a young parent cloud will be poor in H$_2$S and rich in SO$_2$, while a hot core formed from a more evolved parent cloud would be rich in H$_2$S and H$_2$CS. 

\subsection{0D dynamic model} \label{sec_dyn}

In this section we carry out the same study as in the previous section but for the 0D dynamic simulations 0DDLEDC and 0DDEDC described in section \ref{sec_0DD} and table \ref{tab_3}.\\

\begin{figure*}
        \begin{center}
                \includegraphics[scale=0.13]{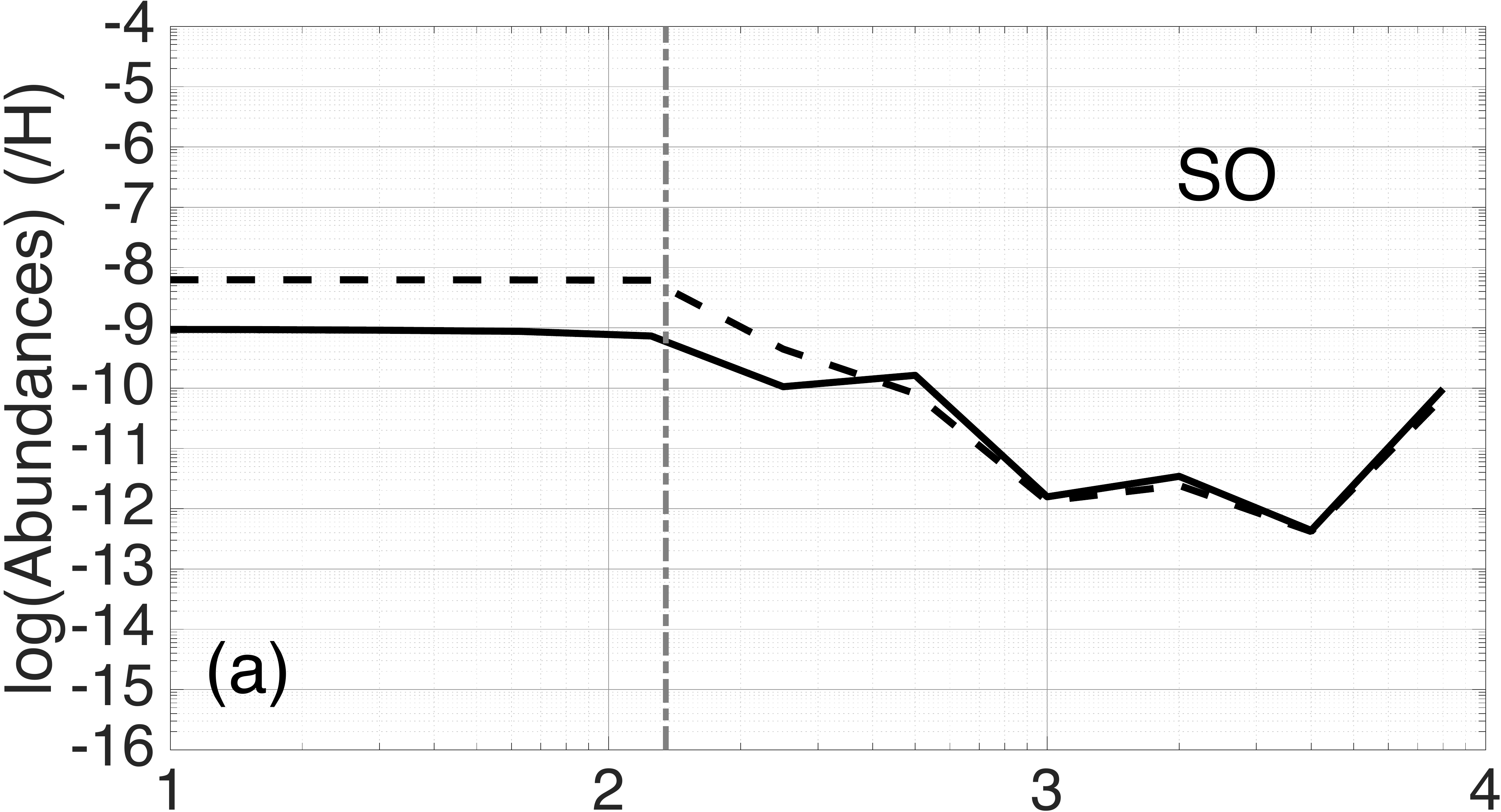}
                \includegraphics[scale=0.13]{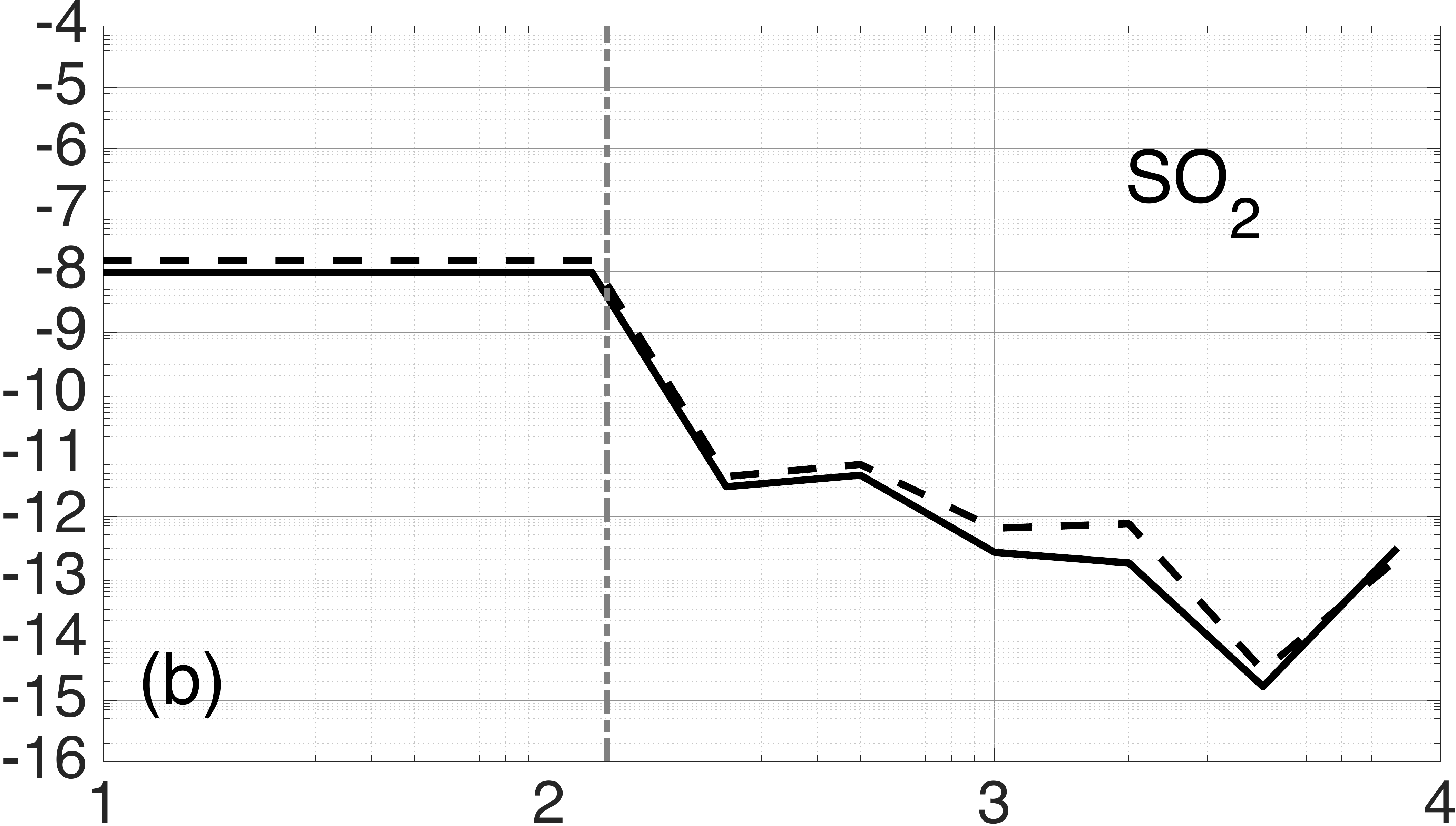}
                \includegraphics[scale=0.13]{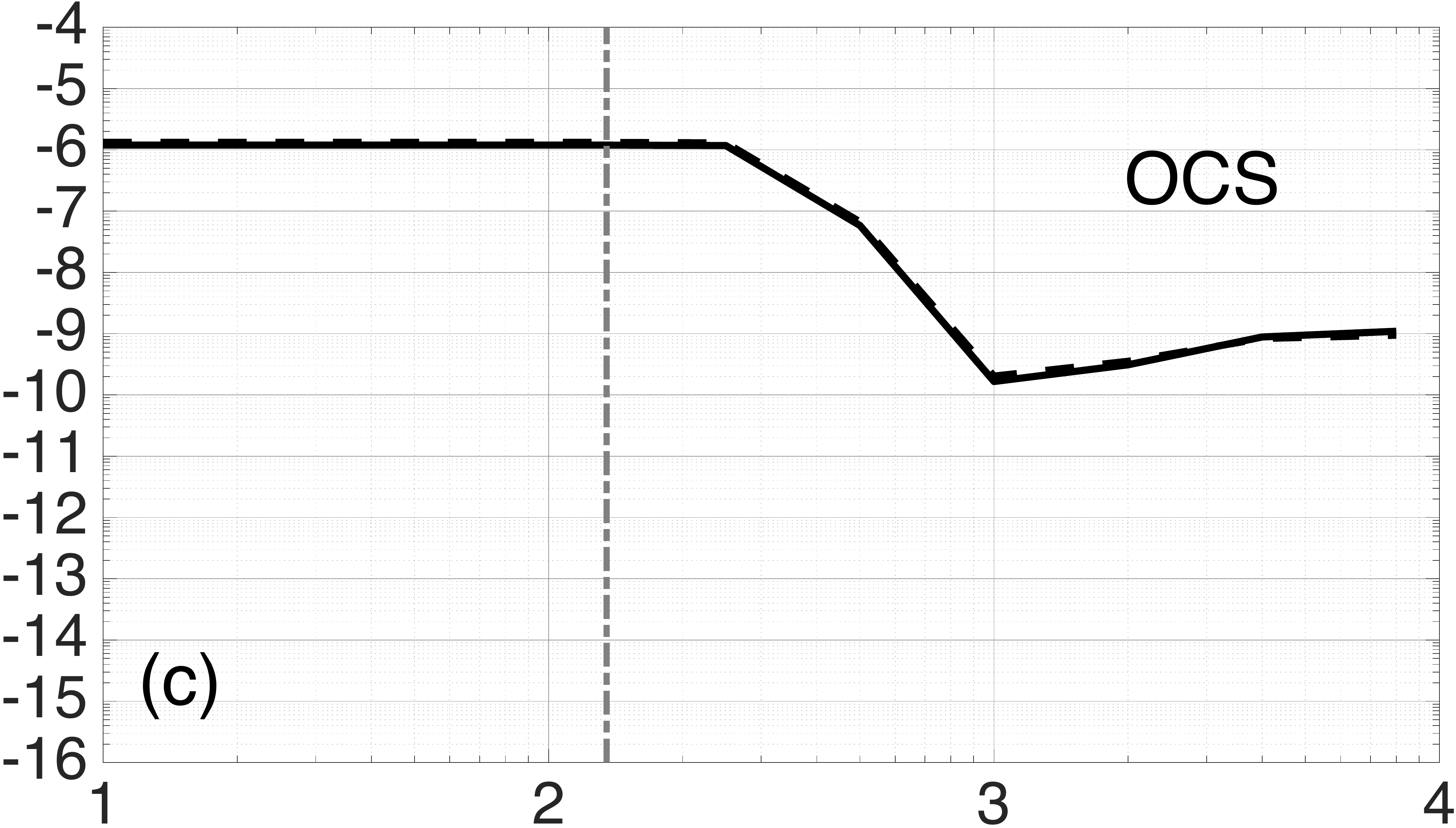}\\
                \includegraphics[scale=0.13]{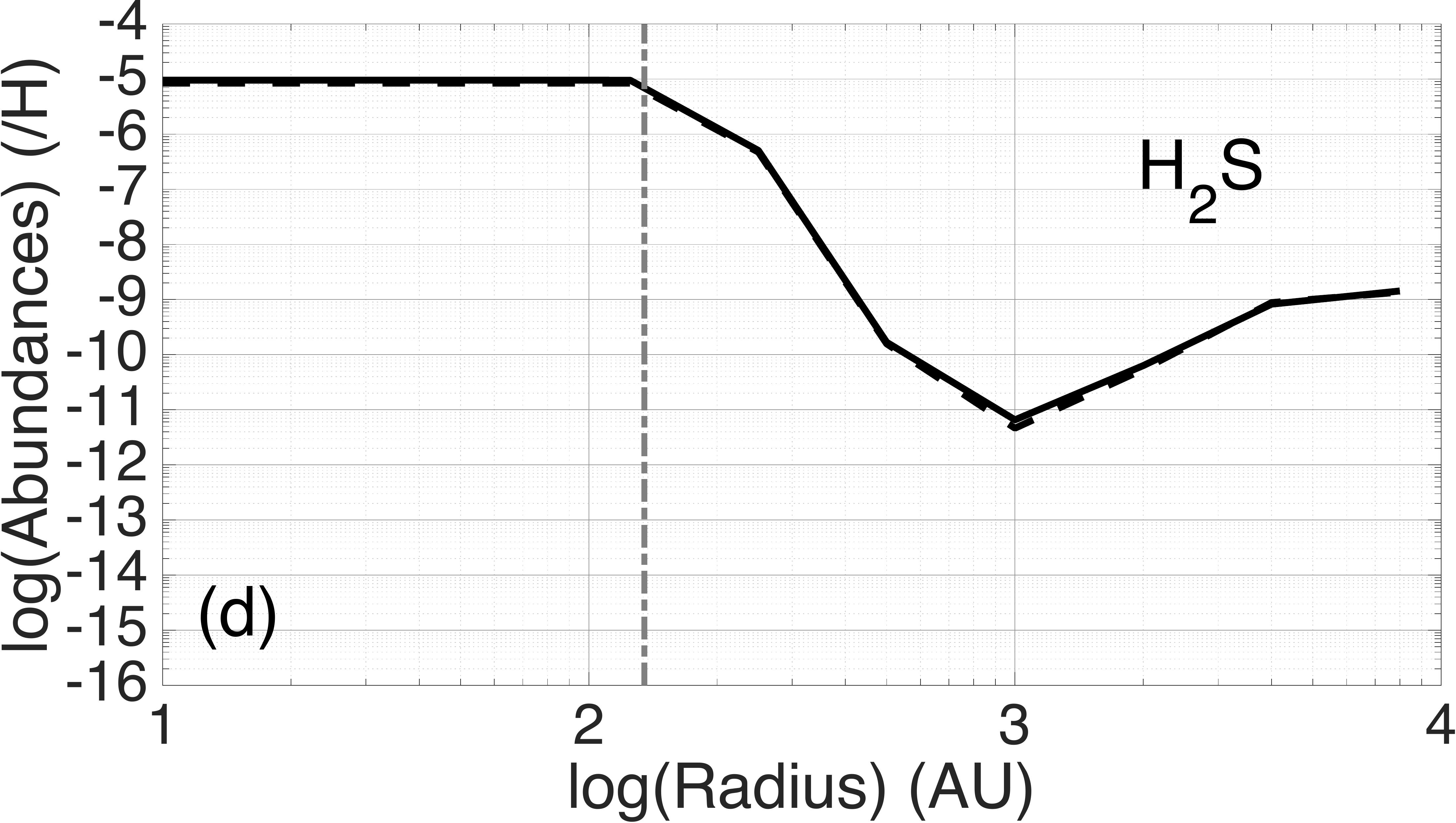}
                \includegraphics[scale=0.13]{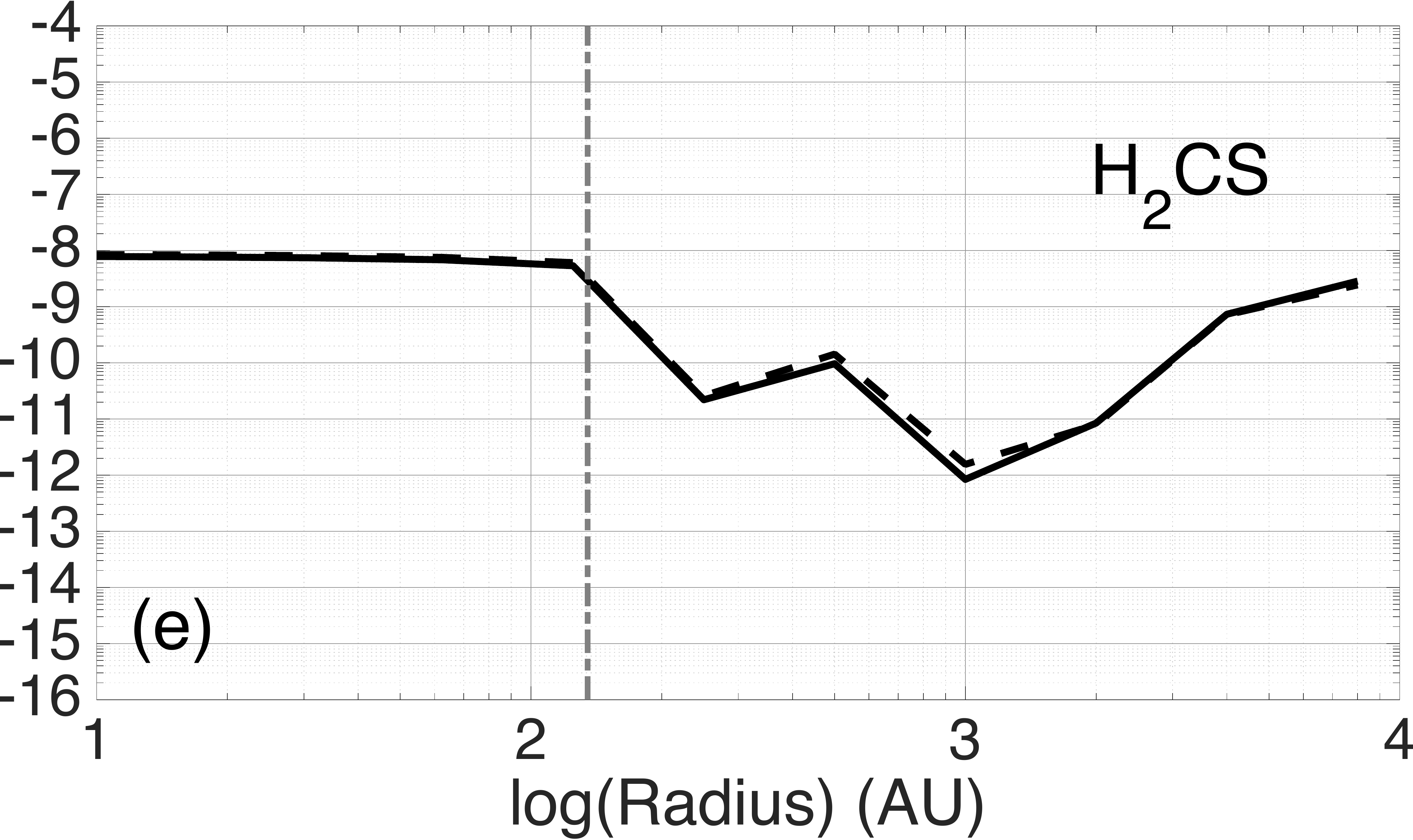}
                \includegraphics[scale=0.13]{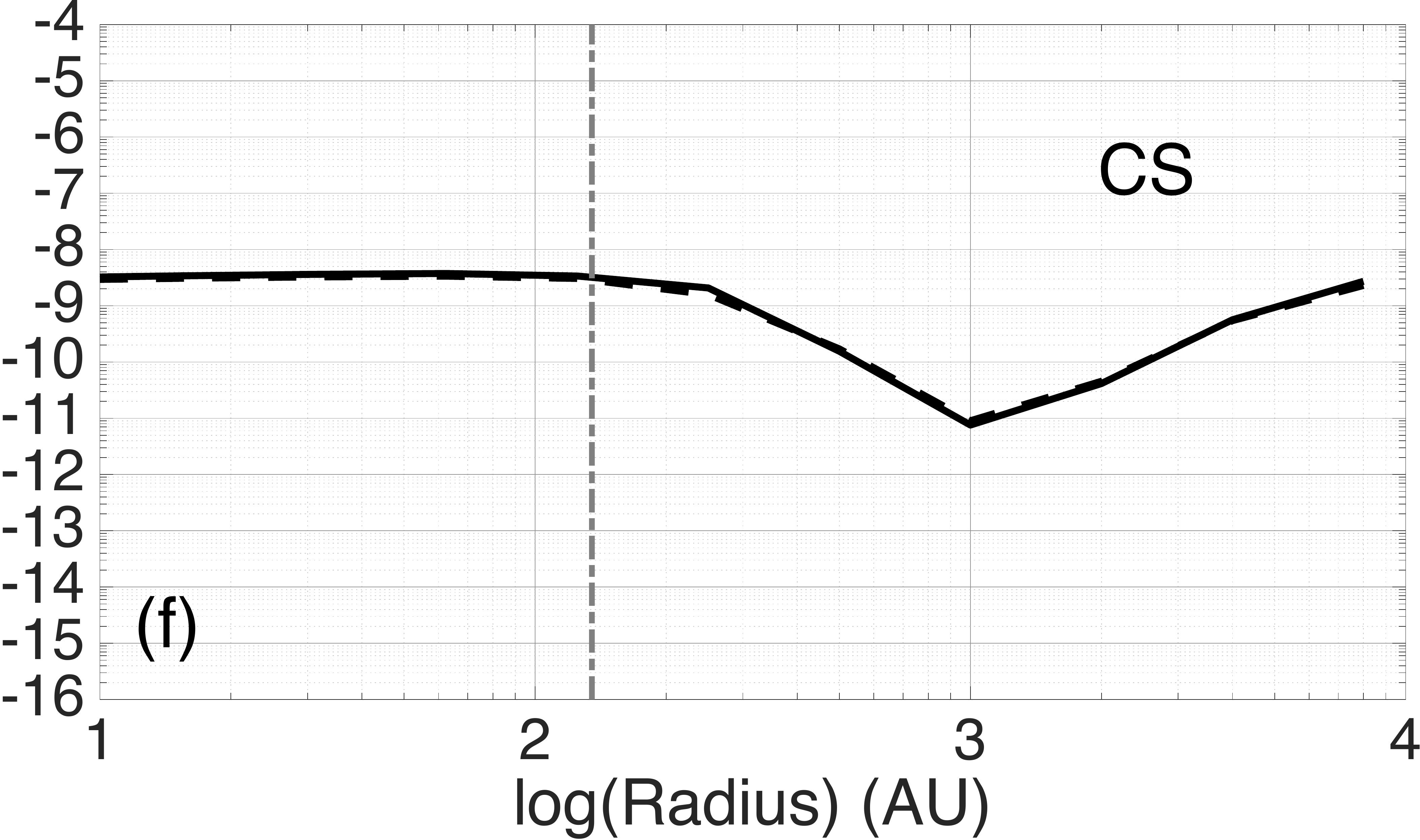}
                \caption{Abundances of SO, SO$_2$ and OCS (top panel) and H$_2$S, H$_2$CS and CS (bottom panel), relative to H as a function of the radius to the star IRAS 16293-2422 according to the modified structure of \citet{Aikawa08}, for the LEDC (solid line) and EDC (dashed line) pre-collapse compositions. The vertical grey dashed line represents the hot core spatial limit $R_{HC}$ = 135 AU, T > 100 K.}
                \label{fig_6}
        \end{center}
\end{figure*}

Figure \ref{fig_6} is the same as figure \ref{fig_5} but for the 0D dynamic simulations. What strikes directly on this figure is that the pre-collapse chemical composition of the parent cloud appears to have little or no effect on the abundances of the considered S-bearing species. Only SO presents significative difference in the hot core (delimited by a vertical grey line in figure \ref{fig_6} at $R_{HC}$ < 135 AU, T > 100 K), and even this difference is not of more than a factor seven (cf figure \ref{fig_6} (a)). A possible explanation of this lack of differences would be that the free-fall time considered in these simulations is long enough for both models to evolve towards the same chemical composition. Indeed, if the initial pre-collapse chemical composition has enough time to evolve in an environment cold enough for species not to evaporate, it will tend towards a state similar to the EDC case, i.e. with an evolved grain surface and bulk chemistry, notably with most of the sulphur transformed into H$_2$S and OCS on the grains. Hence, the hot core chemical composition of both 0DDLEDC and 0DDEDC would be very similar.\\

Moreover, the modification we made on the density profile of the model would tend to accelerate the chemistry and adsorption of species on the grains and therefore reduce the chemical timescale. We expect that by using the original model of \citet{Aikawa08} we would get more differences and thus a more important impact of the pre-collapse chemical composition on the hot core chemistry.

\section{Discussions}

\subsection{About the modification of the density profile of the dynamic model} \label{sec_discdens}

\begin{figure*}
        \begin{center}
                \includegraphics[scale=0.13]{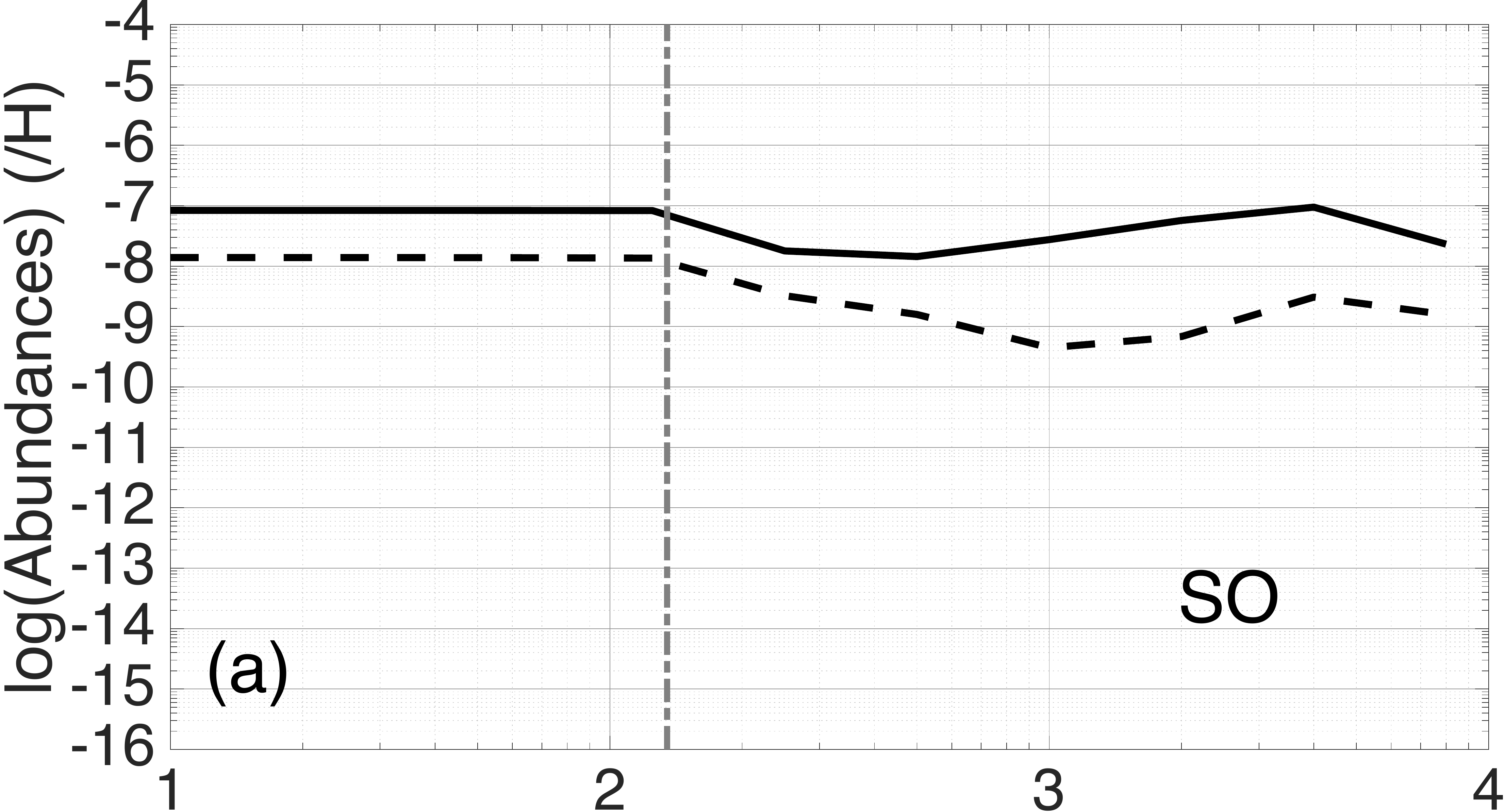}
                \includegraphics[scale=0.135]{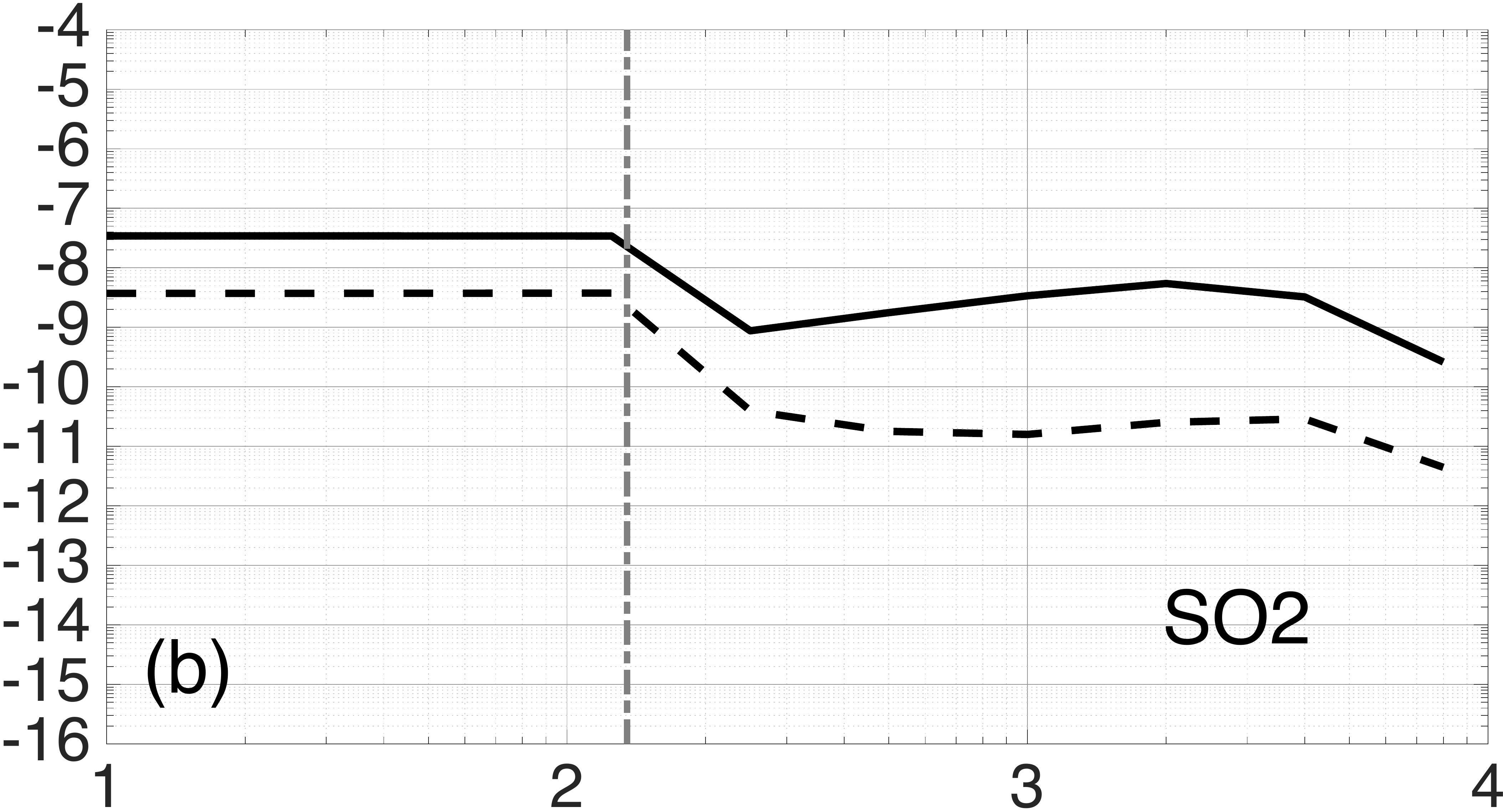}
                \includegraphics[scale=0.13]{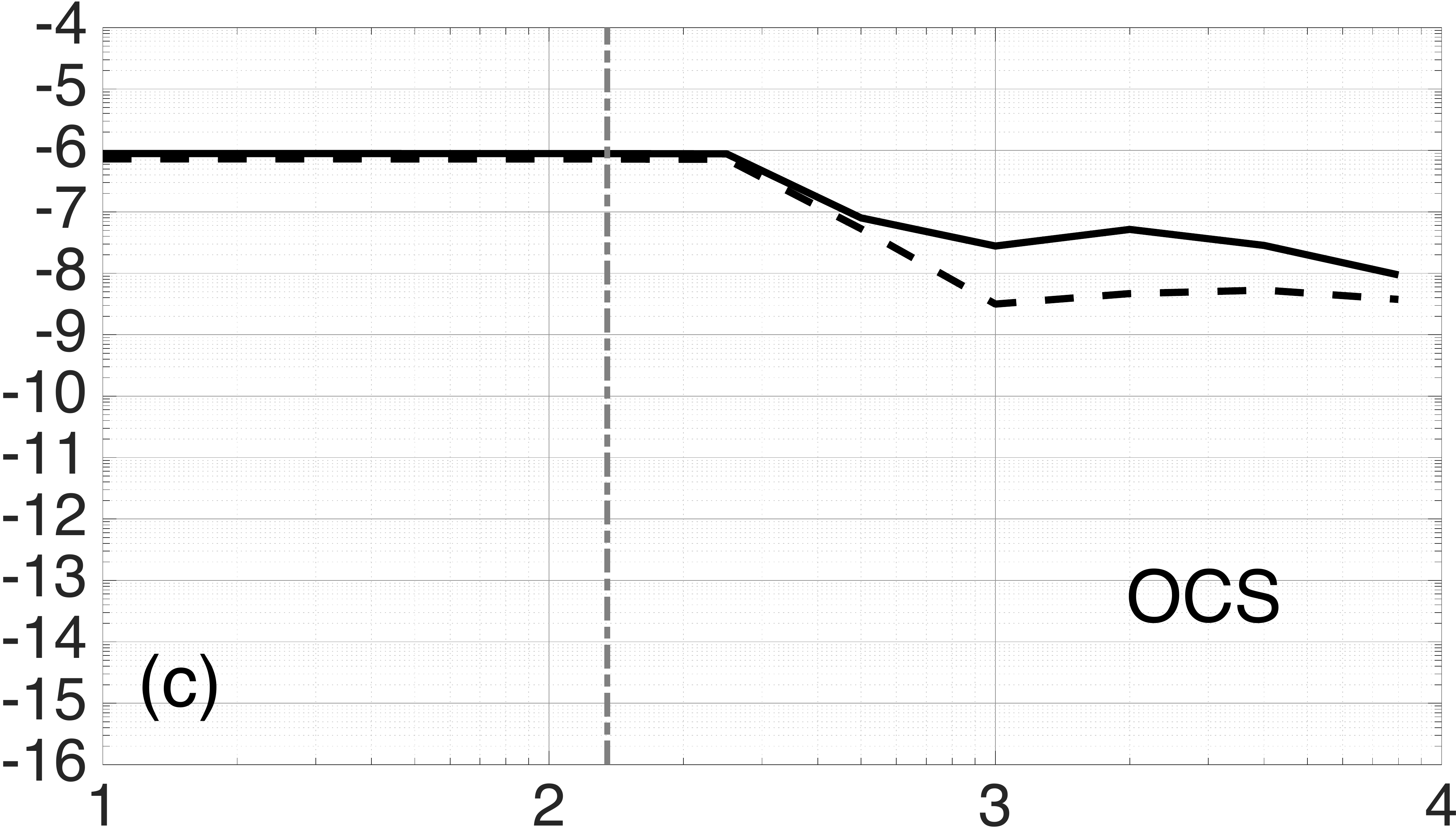}\\
                \includegraphics[scale=0.13]{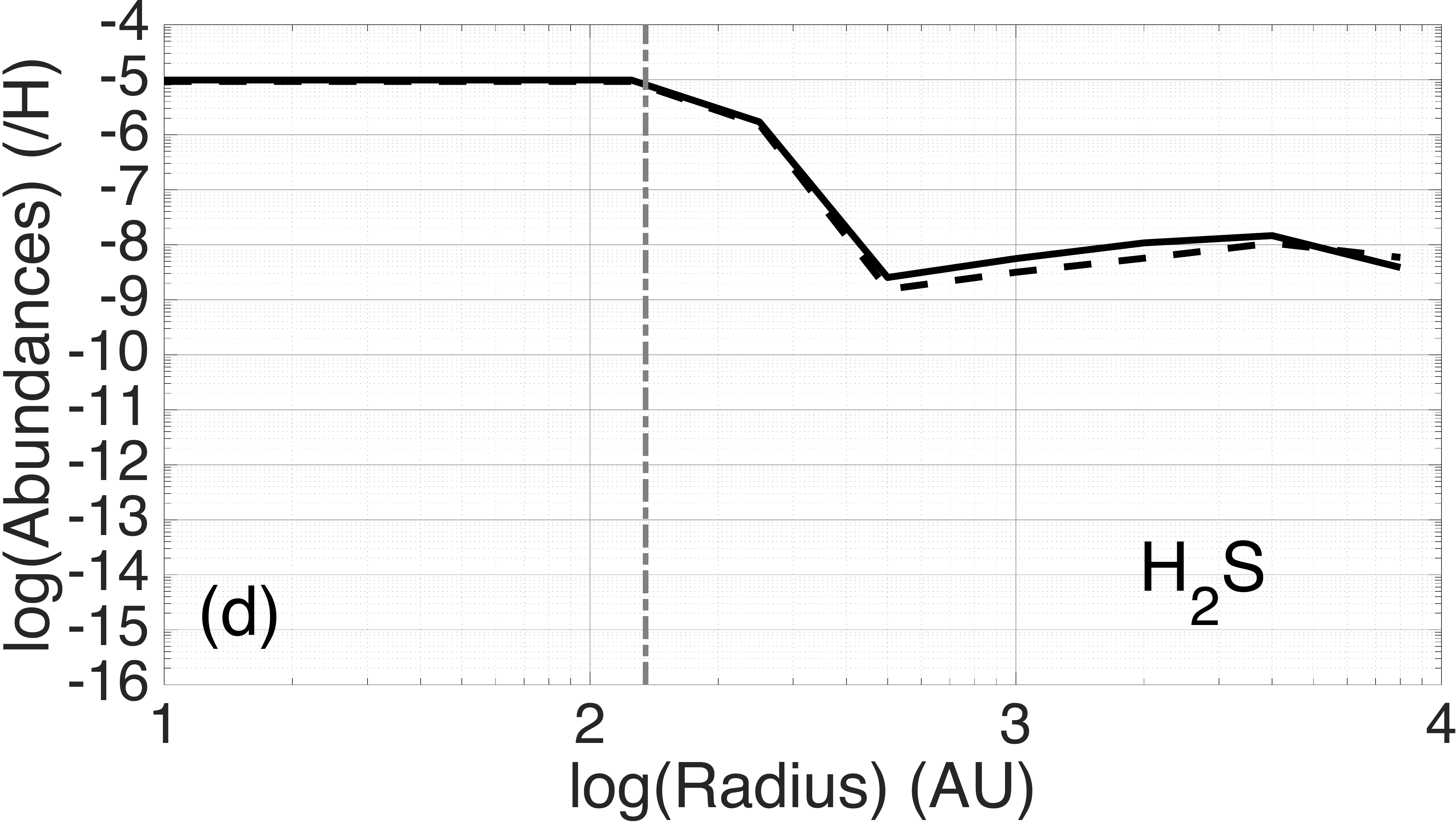}
                \includegraphics[scale=0.13]{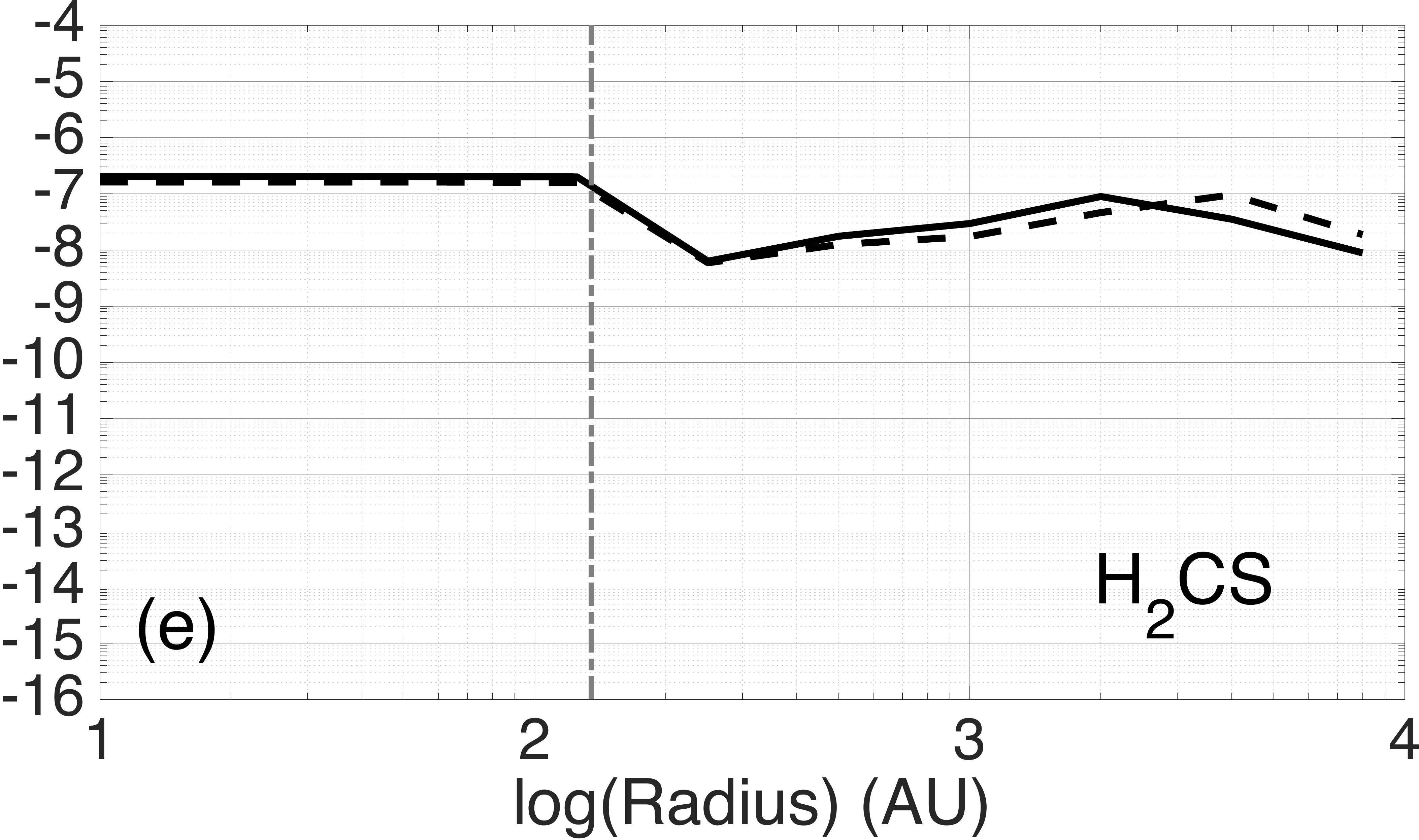}
                \includegraphics[scale=0.13]{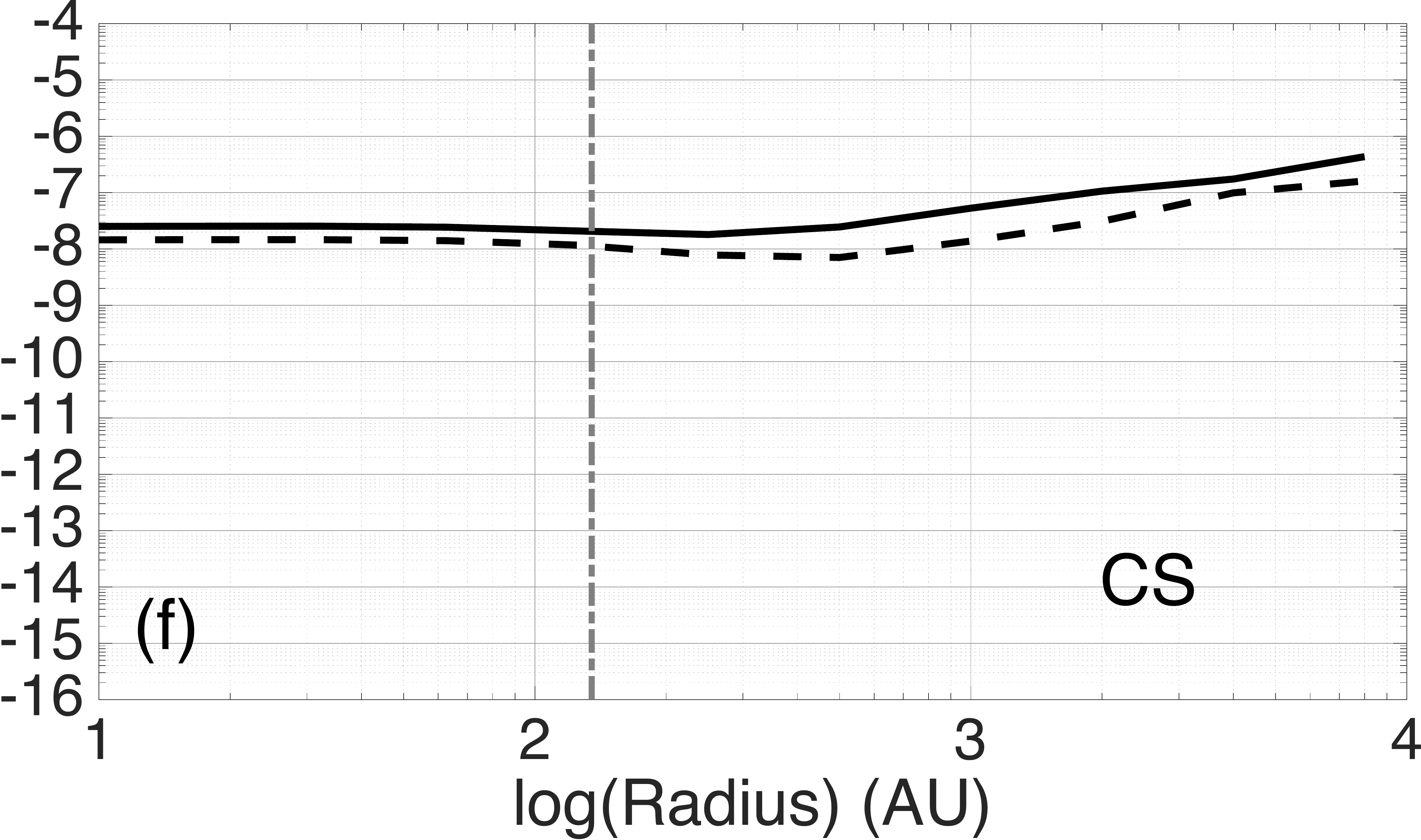}
                \caption{Abundances of SO, SO$_2$ and OCS (top panel) and H$_2$S, H$_2$CS and CS (bottom panel), relative to H as a function of the radius to the star IRAS 16293-2422 according to the original structure of \citet{Aikawa08}, for the LEDC (solid line) and EDC (dashed line) pre-collapse compositions. The vertical grey dashed line represents the hot core spatial limit $R_{HC}$ = 135 AU, T > 100 K.}
                \label{fig_7}
        \end{center}
\end{figure*}

In order to evaluate the impact of the modification of the density radial profile for our 0D dynamic model on sulphur chemistry, we run 0DDLEDC and 0DDEDC cases, but using this time the original physical structure of \citet{Aikawa08}. Figure \ref{fig_7} is therefore the same as figure \ref{fig_6}, but for the models ran with the original density structure of \citet{Aikawa08}. As expected, the results displays larger differences between the LEDC and EDC pre-collapse compositions cases which can be explained by a slower chemistry due to lower density. Moreover, as discussed in \citet{Wakelam14}, lower densities tend to decrease adsorption of species on grains (and consequently depletion), and therefore enhance gas phase chemistry at low temperature, which explains why the abundances of the studied species are higher in the envelope in the original structure case. However the resulting abundances in the hot core do not change drastically (at most a bit more than one order of magnitude) and are even the same for OCS and H$_2$S (cf figure \ref{fig_7} (c) and (d), respectively). As discussed in section \ref{sec_dyn}, this could be explained by a long free-fall time which would let the chemistry enough time for H$_2$S and OCS to accumulate on the grains before thermal depletion in the hot core. More complete studies of the effect of density and free-fall time on the chemistry of a collapsing envelopes and hot cores will be conducted in the future.

\subsection{About H$_2$S and the initial abundance of sulphur}

This study was conducted under the light of the last results we obtained regarding sulphur reservoir in dark clouds of \citet{Vidal17}, namely that depending on the age of the cloud, the reservoir of sulphur could either be atomic sulphur in the gas phase (LEDC case) or H$_2$S and HS in icy grain bulk nearly equally sharing more than 55\% of the total amount of sulphur (EDC case). Another result was that the \textsc{Nautilus} model could reproduce the S-bearing observations in the dark cloud TMC-1 using as initial abundance of sulphur its cosmic one, or three times depleted. Hence, throughout all this paper we present results obtained using the cosmic abundance of sulphur of $1.5\times10^{-5}$ \citep{Jenkins09}. With this initial abundance of sulphur, the H$_2$S abundances obtained in the hot core in the dynamic simulation with the modified as well as the original structure from \citet{Aikawa08} is as high as $10^{-5}$. Such a high abundance of H$_2$S is not consistent with the abundance derived from the observation of IRAS 16293-2422 \citep[$2.7\times10^{-7}$, see][]{Wakelam04b}. On the one hand, to understand why the model overestimates the H$_2$S abundance, we ran all the models presented in this paper with an initial abundance of $5\times10^{-6}$. It first should be noted that it only changes the presented results quantitatively, linearly diminishing the abundances of the studied species by approximately a factor three. Regarding H$_2$S in the dynamical case, this initial depletion of sulphur allows an estimation of its abundance in the hot core slightly overestimated around $3\times10^{-6}$, but which could be considered in accordance with the observations (within a one order of magnitude margin).\\
On the other hand, the overestimation of H$_2$S we find could be due to the efficient formation path due to slowly evaporating CH$_2$SH and CH$_3$OH at 100 K studied in section \ref{sec_0D}. This is in contradiction with previous theoretical and laboratory studies that predict that in high temperature gas phase, the H$_2$S evaporated from grain ices is preferentially destroyed to form SO and SO$_2$, or molecules with two S atoms such as H$_2$S$_2$ or HS$_2$ \citep{Charnley97,Wakelam04,Druard12,Esplugues14,Martin16}. Hence, this result could be a hint of missing efficient destruction gas phase reactions in the H$_2$S chemistry in our network.

\subsection{About the sensitivity to the type of simulation}

\begin{table}
\caption{Comparison of the abundances obtained in the LEDC case. $a(b)$ stands for $a\times10^b$}
	\begin{center}
		\begin{tabular}{c c c}
		\hline
		\hline
		Species & $(n_i/n_H)_{\text{1DSLEDC}}$  & $(n_i/n_H)_{\text{0DDLEDC}}$\\
		\hline
   		\multicolumn{3}{c}{R = 50 AU (Hot core)} \\
   		\hline
		SO & 5.7(-8) & 9.0(-10)\\
		SO$_2$ & 1.4(-5) & 9.5(-9)\\
		OCS & 3.0(-7) & 1.2(-6)\\
		H$_2$S & 1.2(-11) & 9.5(-6)\\
		H$_2$CS & 7.4(-10) & 7.6(-9)\\
		CS & 1.2(-12) & 3.3(-9)\\		
		\hline
		\multicolumn{3}{c}{R = 500 AU (Envelope)} \\
		\hline
		SO & 2.3(-14) & 1.6(-10)\\
		SO$_2$ & 1.2(-13) & 4.7(-12)\\
		OCS & 2.2(-12) & 5.8(-8)\\
		H$_2$S & 1.1(-13) & 1.7(-10)\\
		H$_2$CS & 2.7(-14) & 1.4(-10)\\
		CS & 1.2(-13) & 1.7(-10)\\	
		\hline			
 		\end{tabular}
	\end{center}
  	\label{tab_5}
\end{table}

One goal of this article was to highlight the differences that can appear when using two types of simulations to model the same hot core. In order to do so we used a 1D static model and a 0D dynamic model of IRAS 16293-2422. Our results show that our 1D model favors a hot core sulphur chemistry dominated by SO$_2$ and SO in the LEDC case and H$_2$S in the EDC case while our 0D dynamic model displays in both cases high abundances of H$_2$S and OCS and low abundances of SO$_2$ and SO. Table \ref{tab_5} displays the abundances obtained for both types of models in the LEDC case, at 50 AU (in the hot core) and 500 AU (in the envelope). At the light of the difference in abundance that exists for a given species between the two type of models, one can easily conclude on the sensitivity to the type of simulation to use to compute the chemistry of a hot core and its collapsing envelope. Especially in the hot core, these differences can reach as much as six orders of magnitude, rendering critical the choice of model used to compare results with possible observations or related works.

\subsection{About the importance of the pre-collapse chemical composition}

The observations of S-bearing species in hot cores are still a puzzling issue, since a large variety of sulphur compositions have been observed towards different hot cores and therefore no global trend has yet been found \citep[see figure 5 of][and references therein]{Woods15}. However, a given set of hot cores can present similar sulphur compositions \citep[see for example][]{Minh16}, which would suggest similar evolutionary stages. In this paper we investigated the importance of the pre-collapse chemical composition on the hot core chemistry of S-bearing species. Our results on 0D and 1D simulations (section \ref{sec_0D} and \ref{sec_1D}) support that, given the fast evolution of sulphur chemistry in the parent cold clouds, the pre-collapse chemical composition is a critical parameter for hot core simulations. This could partially explains the absence of global trend for sulphur compositions in observed hot cores from different parent clouds, as well as supports the fact that for parent clouds collapsing at a similar ages and physical environments, hot cores can have similar sulphur compositions. However, the study of our dynamical simulation (section \ref{sec_dyn}) raises the question of the role of the free-fall time on sulphur evolution, which would also explains similar composition for different hot cores, especially those where H$_2$S and OCS are found more abundant than the other species \citep[see for example][]{Herpin09}.

\section{Conclusions}

In this paper we aimed to take a comprehensive look at the chemistry of sulphur in hot cores. In order to do so, we first conducted an extensive study with simple 0D models of the chemistry of the main S-bearing species observed towards hot cores, namely SO, SO$_2$, OCS, H$_2$S, H$_2$CS and CS. We then presented and compared the results from two types of simulations (1D static and 0D dynamic), in order to highlight the sensitivity of chemistry to the choice of model used in astrochemical studies.\\

Our 0D extensive study revealed four main results:

\begin{enumerate}
\item The total amount of reactive oxygen in the gas phase is critically depending on the pre-collapse composition of the hot core (cf table \ref{tab_4}) as well as the temperature;\\
\item Sulphur chemistry in hot and dense gas depends also highly on the pre-collapse composition, mainly because of its impact on reactive atomic oxygen, carbon and hydrogen, which all participate actively in most of the sulphur chemistry in such environment;\\
\item Sulphur chemistry in hot and dense gas depends highly on the temperature, partly because it is directly and indirectly linked with hydrocarbon evaporated from grain ices, and their main destruction products CH$_2$ and CH$_3$; \\
\item We found efficient paths of formation of gas phase H$_2$S that could be responsible for its overestimation in most of our hot core results. Studies of the gas phase chemistry of this species need to be continued to ensure the relevance of our network.
\end{enumerate}

Our study of the 1D static and 0D dynamic models led to the following conclusions:

\begin{enumerate}
\item The pre-collapse chemical composition of the parent cloud is a key parameter for 1D static simulations of sulphur chemistry in hot cores. Indeed, the computed abundances showed that it can imply differences up to six orders of magnitude for a given species in the hot core. However, the pre-collapse composition appears to only have a small impact on the chemical composition of the envelope. Finally our 1D model shows that a hot core that was formed from a young parent cloud will be poor in H$_2$S and rich in SO$_2$, while a hot core formed from a more evolved parent cloud would be rich in H$_2$S and H$_2$CS.\\
\item The 0D dynamic simulations conducted in this paper revealed only small differences between the results of the less evolved and the evolved pre-collapse chemical composition, showing only a weak dependence of the hot core sulphur chemistry on the pre-collapse composition. Indeed, for both cases the model predicts high abundances of H$_2$S and OCS and low abundances of SO$_2$ and SO. However, this result is thought to be due to the rather long free-fall time used in our model, which would let enough time for sulphur to be adsorbed on grains and to form mainly H$_2$S and OCS in the envelope before thermal depletion. We expect that for a shorter free-fall time the differences between the two pre-collapse composition cases would be larger. Future work will focus on this importance of the free-fall time, since it could, along with the pre-collapse chemical composition, explain the large variety of abundances of S-bearing species observed in hot cores.
\end{enumerate}

The comparison between the 1D static and 0D dynamic models displayed large differences on the computed abundances that can go as high as six orders of magnitude in the hot core. This result highlights the sensitivity to the choice of simulations in astrochemical study, especially when comparing results with observations, or with results from other paper.

\section*{Acknowledgements}
This work has been founded by the European Research Council (Starting Grant 3DICE, grant agreement 336474). The authors are also grateful to the CNRS program "Physique et Chimie du Milieu Interstellaire" (PCMI) for partial funding of their work.




\bibliographystyle{mnras}
\bibliography{bibliography} 




\bsp	
\label{lastpage}
\end{document}